\documentclass[a4paper,12pt]{article}
\usepackage{jheppub,esint,psfrag}
\usepackage[utf8]{inputenc}

\usepackage{graphicx}
\usepackage{caption}
\usepackage{subcaption}
\usepackage{placeins}
\usepackage{amsmath}
\usepackage{amsfonts}
\usepackage{mathtools}

\usepackage{epsfig,amssymb,amsmath,psfrag,rotate,color,wasysym,xcolor}
\usepackage[bbgreekl]{mathbbol}

\allowdisplaybreaks 
\newcommand{\insertfig}[2]{\includegraphics[width=#1cm]{#2}}

\DeclareSymbolFontAlphabet{\mathbbm}{bbold}
\DeclareSymbolFontAlphabet{\mathbb}{AMSb}%

\def\XXint#1#2#3{{\setbox0=\hbox{$#1{#2#3}{\int}$ }
\vcenter{\hbox{$#2#3$ }}\kern-.6\wd0}}

\def \be  {\begin{equation}}
\def \ee  {\end{equation}}
\def \ba  {\begin{eqnarray}}
\def \ea  {\end{eqnarray}}
\def \baa {\begin{eqnarray*}}
\def \eaa {\end{eqnarray*}}
\newcommand{\ep}{\varepsilon}
\def \lab #1 {\label{#1}}

\newcommand\re[1]{(\ref{#1})}
\def\d{\hbox{{d}\kern-.20em\hbox{l}}}

\def \matrix #1 {\left(\begin{array}{cc} #1 \end{array}\right)}

\def \tr {\mathop{\rm tr}\nolimits}

\def \e  {\mathop{\rm e}\nolimits}

\newcommand \ket [1] {|{#1}\rangle}
\newcommand \bra [1] {\langle {#1}|}
\newcommand{\bit}[1]{\mbox{\boldmath$#1$}}
\def\1{\hbox{{1}\kern-.25em\hbox{l}}}

\newcommand{\ft}[2]{{\textstyle\frac{#1}{#2}}}

\makeatletter

\input pdf-trans
\newbox\qbox
\def\usecolor#1{\csname\string\color@#1\endcsname\space}
\newcommand\bordercolor[1]{\colsplit{1}{#1}}
\newcommand\fillcolor[1]{\colsplit{0}{#1}}
\newcommand\outline[1]{\leavevmode%
  \def\maltext{#1}%
  \setbox\qbox=\hbox{\maltext}%
  \boxgs{Q q 2 Tr \thickness\space w \fillcol\space \bordercol\space}{}%
  \copy\qbox%
}
\makeatother
\newcommand\colsplit[2]{\colorlet{tmpcolor}{#2}\edef\tmp{\usecolor{tmpcolor}}%
  \def\tmpB{}\expandafter\colsplithelp\tmp\relax%
  \ifnum0=#1\relax\edef\fillcol{\tmpB}\else\edef\bordercol{\tmpC}\fi}
\def\colsplithelp#1#2 #3\relax{%
  \edef\tmpB{\tmpB#1#2 }%
  \ifnum `#1>`9\relax\def\tmpC{#3}\else\colsplithelp#3\relax\fi
}
\bordercolor{black}
\fillcolor{white}
\def\thickness{.3}

\def\1{\mathbbm{1}}

\title{Off-shell form factor: factorization is violated}

\author[a]{A.V.~Belitsky,}
\author[b]{V.A. Smirnov}
\affiliation[a]{Department of Physics, Arizona State University, Tempe, AZ 85287-1504, USA}  
\affiliation[b]{Skobeltsyn Institute of Nuclear Physics, Moscow State University, 119992 Moscow, Russia\\
Moscow Center for Fundamental and Applied Mathematics, 119992 Moscow, Russia}
    
\abstract
{We study the Sudakov form factor on the Coulomb branch of $\mathcal{N} = 4$ sYM which endows only external states 
with masses, and implies that the former is off-shell in the traditional sense. Our consideration is performed at three-loop 
order in the near mass-shell limit. We use a combination of tools to perform required calculations centered around the Method 
of Regions as the main go-to formalism for the asymptotic expansion of emerging parametric Feynman integrals. Explicit 
separation of quantum loops in terms of hard, collinear, and ultrasoft modes allows us to explore the factorization properties of 
this infrared-sensitive quantity. While the hard region is cleanly separated from the rest, the ultrasoft-collinear modes remain 
intertwined. We exhibit effects of factorization violation explicitly in the momentum space making use of the infrared power 
counting.}

\begin{document}

\maketitle
\flushbottom
\setcounter{footnote} 0

\section{Introduction}

The Sudakov form factor \cite{Sudakov:1954sw} is the simplest infrared-sensitive matrix element, which describes the production of a 
colored particle pair by a gauge-invariant operator source from the vacuum. It thus represents the simplest laboratory to explore gauge 
dynamics. To use the full predictive power of perturbative quantum field theory, one must establish a clear separation of physics at different 
space-time scales involved. Intuitively, this should occur due to the immunity of the short-distance, aka hard, scattering to long wave-length, 
aka infrared (IR), (nearly) real emissions. The latter can affect the former only incoherently. IR components per se can be classified 
according to the scaling pattern of their momenta with respect to a soft scale intrinsic to processes with gauge bosons. If the latter are 
predominantly emitted in the direction of their parent, they define the collinear regime, while if all components of gauge momenta 
are small, they correspond to the (ultra)soft domain. Classically, one anticipates that (ultra)soft modes feel only the overall 
color charge of the fast propagating jet of collinear emissions and, similarly to the hard factorization alluded to above, influences 
it only incoherently. These general arguments require proof. Proofs are observable-specific and are not universal. They are not 
warranted.

Typically, factorization theorems rely on Landau equations \cite{Landau:1959fi,Coleman:1965xm} and establish pinch surfaces of 
amplitudes/cross sections where propagators go on-shell and blow up. These allow one to devise their decomposition in terms of 
various IR components. One of the central elements in these considerations is the ability to use eikonal, or Grammer-Yennie, 
approximation. Then, a precise form for various incoherent building blocks can be established in terms of matrix elements of certain 
nonlocal operators. Over the years, a variety of techniques has been used for these purposes: (i) original QCD-field formulation 
\cite{Sen:1981sd,Sen:1982bt,Collins:1989gx,Sterman:2002qn,Aybat:2006mz,Dixon:2008gr,Becher:2009qa,Agarwal:2021ais}, 
(ii) soft-collinear effective theory that introduces individual modes for each IR regime \cite{Bauer:2000yr,Beneke:2002ph,Becher:2014oda}, 
and (iii) effective field theory \cite{Feige:2013zla,Feige:2014wja}.

There are processes where the above factorization theorems are known to be broken. One of the most well-studied 
cases corresponds to the situation that receives leading power-counting contributions from the so-called Glauber gluons 
\cite{Collins:1981ta,Bodwin:1981fv,Collins:1983ju,Rothstein:2016bsq,Schwartz:2017nmr}. The latter are distinguished from 
other IR modes by the fact that they possess transverse (with respect to some collinear directions) momentum which is much 
larger than their energy/longitudinal components. These induce interactions between initial and final states and are known to 
invalidate strict factorization of amplitudes \cite{Collins:1983ju,Collins:1988ig}. They can vanish, however, at a cross-section level 
if observables are sufficiently inclusive \cite{Collins:1983ju,Collins:1988ig,Aybat:2008ct,Schwartz:2018obd}, or ruin them for non-inclusive 
ones altogether \cite{Forshaw:2006fk,Becher:2021zkk,Boer:2024hzh}. The latter effects manifest themselves through the appearance of 
super-leading logarithms at high perturbative orders. Apparently, a different violation of strict factorization takes place in exceptional 
kinematics when some of the initial and final state momenta become collinear \cite{Catani:2011st,Forshaw:2012bi,Schwartz:2017nmr}. 
However, its physical origin is related to the emergence of super-leading logarithms \cite{Forshaw:2012bi}.

In this paper, we address hard-ultrasoft-collinear factorization of the off-shell Sudakov form factor. As compared to the aforementioned 
works by other authors, we will rely on the Method of Regions \cite{Beneke:1997zp} (see also Refs.\ \cite{Smirnov:2002pj,Smirnov:2012gma}
and \cite{Smirnov:2021dkb} for extensive and concise reviews, respectively) as a tool for the classification and extraction of IR physics from 
perturbative graphs order by order in perturbation theory. This framework has the advantage of providing a precise definition for leading 
momentum regions in IR power counting. Having established their integral form, one can then devise matrix elements of operators that 
reproduce them in the perturbative expansion. This task was accomplished at two loops in Ref.\ \cite{Belitsky:2024yag}. Presently, we push 
it to the next perturbative order, i.e., three loops. 

In this work, the off-shell Sudakov form factor is studied in a `toy' gauge theory, the maximally supersymmetric Yang-Mills theory, aka $\mathcal{N}=4$
sYM, on the Coulomb branch. The reason for this is twofold. First, this model provides a simple way to construct a gauge-invariant off-shell 
generalization of on-shell amplitudes. Second, the off-shell Sudakov form factor in the near mass-shell limit possesses an exact all-order form, 
with the off-shellness serving as an IR regulator, so it allows one to study (ultra)soft-collinear physics unobstructed by regularization issues.

Our subsequent presentation is organized as follows. In the next section, we define the off-shell Sudakov form factor and introduce
IR power counting for integration momentum in one-loop graphs. Next, we recall our results at two-loop order before turning to 
the three-loop consideration within the Method of Regions. In Sect.\ \ref{HardSect}, we factor out the hard region from the ultrasoft-collinear
modes and discuss complications in separating the latter two. In Sect.\ \ref{UScolSect}, we pinpoint the origin of the obstruction. Finally, we
conclude.

\section{Sudakov form factor and IR power counting}

\begin{figure}[t]
\begin{center}
\mbox{
\begin{picture}(0,345)(230,0)
\put(0,0){\insertfig{16}{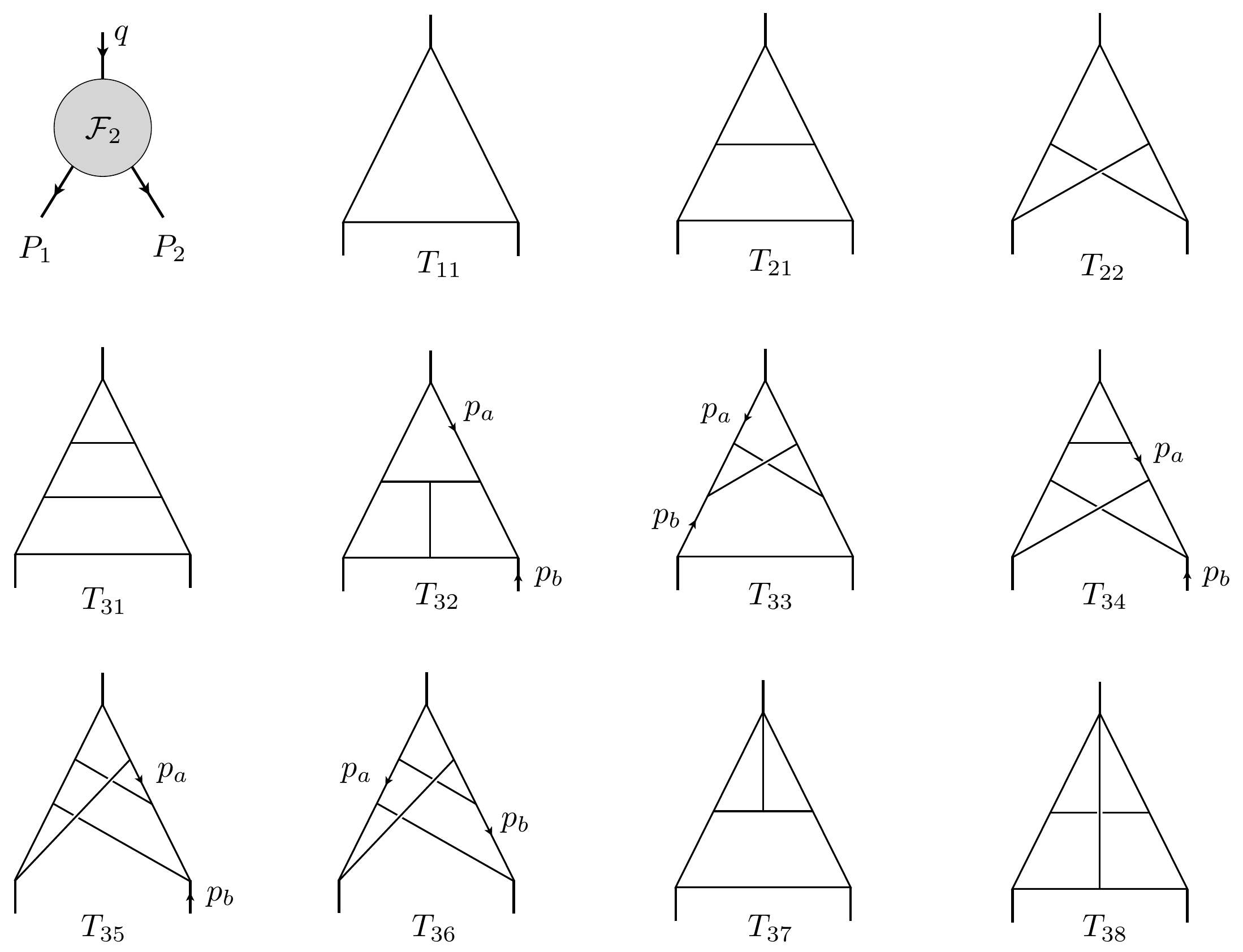}}
\end{picture}
}
\end{center}
\caption{\label{123IntegralsFig} Generic representation for the Sudakov form factor (top leftmost figure) and contributing momentum
integrals to three-loop order. Arrows on perturbative graphs imply the presence of an irreducible numerator $(p_a + p_b)^2$.}
\end{figure}

The Sudakov form factor $\mathcal{F}_2 $ in $\mathcal{N} = 4$ sYM is defined as a transition matrix element 
\begin{align}
\label{SudakovDef}
(2 \pi)^4 \delta^{(4)} (q - P_1 - P_2) \, \mathcal{F}_2 
=
\int d^4 x \e^{- i q \cdot x} \bra{P_1, P_2} \mathcal{O} (x) \ket{0}
\end{align}
of the half-BPS
operator built from the sextet of scalar fields $\phi_{AB}$ of the model,
\begin{align}
\mathcal{O} (x) = \frac{1}{2N_c} \tr_{\rm adj} \phi_{12}^2 (x)
\, .
\end{align}
This is the lowest component of the stress-tensor multiplet and, thus, it does not renormalize. On the Coulomb branch of the theory, which can be obtained 
by a generalized dimensional reduction from the six-dimensional $\mathcal{N} = (1,1)$ sYM \cite{Dennen:2009vk,Bern:2010qa,Belitsky:2024rwv}, 
one can furnish only external legs $P_{1,2}$ with masses \cite{Caron-Huot:2021usw,Belitsky:2024rwv}, while keeping all other lines massless. We 
will interpret the former as virtualities. Moreover, we will pass to the Euclidean kinematics and choose all invariants to be space-like, i.e., 
\begin{align}
q^2 = - Q^2 \, , \qquad P_i^2 = - m^2 \, , \qquad Q^2 \gg m^2 \to 0
\, .
\end{align}
Here, we also imposed a strong ordering on $m$ and $Q$ since we will be only interested in the near mass-shell limit. Keeping $m$ small but 
non-vanishing will serve as a regulator of IR divergences. This setup differs from the on-shell case \cite{Jackiw:1968zz}, where one needs to 
introduce an additional regulator to tame IR singularities. This is traditionally done by continuing the dimension of space-time away from four 
\cite{vanNeerven:1985ja,Gehrmann:2011xn} or by means of a Higgs mechanism \cite{Alday:2009zm}. In the following consideration, we will
set $Q^2 = 1$ without loss of generality since up an overall scale $\mathcal{F}_2$ depends on the dimensionless variable $t = m^2/Q^2$, such
that with this choice the off-shell Sudakov form factor is a function of the virtuality $t$ only. We will use $t$ as the IR scale throughout our 
consideration.

In this paper, we will study \re{SudakovDef} within perturbation theory. More specifically, we will calculate it to the highest currently achievable 
order, i.e., three loops. The expansion 
\begin{align}
\mathcal{F}_2 = \sum_{\ell} g^{2 \ell} \mathcal{F}_2^{(\ell)}
\end{align}
runs in the 't Hooft coupling 
\begin{align}
\label{DtHooft}
g^2 \equiv \e^{-\ep \gamma_{\rm\scriptscriptstyle E}} \frac{g_{\rm\scriptscriptstyle YM}^2 N_c}{(4 \pi)^{D/2}}
\, .
\end{align}
For the reasons that will become clear very shortly, we define it in $D = 4 - 2 \varepsilon$ dimensions. 

The tree-level form factor is normalized to one, $\mathcal{F}_2^{(0)} = 1$. The next three orders $\ell = 1,2,3$ were calculated in 
Refs.\ \cite{Belitsky:2022itf,Belitsky:2023ssv} employing the method of canonical differential equations \cite{Henn:2013pwa} for
basis momentum integrals at each perturbative order that were constructed \cite{Gehrmann:2011xn} with the unitarity-cut sewing 
procedure of Refs.\ \cite{Bern:1994cg,Bern:1994zx}. What was found as a result of that consideration is that the off-shell Sudakov
form factor exponentiates
\begin{align}
\label{ExactOffSud}
\log \mathcal{F}_2 = - \ft12 \Gamma_{\rm oct} (g) \log^2 t -  D (g)
\, ,
\end{align}
for the IR double logarithmic part as well as its finite part $D (g)$ and the first three terms in the 't Hooft expansion coincide with
the  octagon anomalous dimensions $\Gamma_{\rm oct}$ and the overall normalization of the so-called octagon form factor 
\cite{Belitsky:2019fan}
\begin{align}
\label{ADs}
\Gamma_{\rm oct} (g)
= \frac{2}{\pi^2} \log \cosh(2\pi g)
\, , \qquad
D (g) = \frac{1}{4} \log\frac{\sinh (4 \pi g)}{4 \pi g}
\, .
\end{align}
This was yet another manifestation of an observation that a four-point correlator of heavy half-BPS operators can be interpreted as a
scattering amplitude of massive W-bosons \cite{Caron-Huot:2021usw}. With this IR-sensitive observable possessing such a simple 
all-order form, it is only natural to test its factorization properties. 

At one loop, it receives a single contribution\footnote{We follow the nomenclature of Ref.\ \cite{Belitsky:2024yag} and thus keep the 
same sign assignments.} 
\begin{align}
\label{F21}
\mathcal{F}_2^{(1)} = - 2 T_{11}
\end{align}
from a triangle Feynman integral, see Fig.\ \ref{123IntegralsFig}, built from the products of the scalar propagators $D (k) = [k^2 + i 0]^{-1}$
\begin{align}
\label{T11}
T_{11}
&
=
- \int_k D (k) D (k - P_1) D (k + P_2)
\, ,
\end{align}
over the dimensionally-regularized measure
\begin{align}
\label{Dmeasure}
\int_k \equiv 
\e^{\ep \gamma_{\rm\scriptscriptstyle E}} \mu^{2 \ep} \int \frac{d^D k}{i \pi^{D/2}}
\, .
\end{align}
We will set $\mu = 1$ in what follows. To detect the pinch surfaces as $t \to 0$, we employ the Landau criterium that the inverse propagators 
have to vanish or be proportional to external momenta \cite{Landau:1959fi}. To discuss it in a transparent fashion, it is instructive to use the Sudakov 
decomposition for the loop momentum
\begin{align}
k = (k_+, k_-, k_\perp) = k_+ p_1 + k_- p_2 + k_\perp \, ,
\end{align}
in terms of two light-like vectors $p_1$ and $p_2$, with $p_1 \cdot p_2 = - \ft12$, which encode predominant directions of the nearly on-shell 
external legs 
\begin{align}
P_1 = p_1 + t p_2
\, , \qquad
P_2 = p_2 + t p_1
\, ,
\end{align}
respectively. Here, without loss of generality we set $P_{1,+} = P_{2, -} = 1$. In this manner, one uncovers the IR scaling for the loop 
momentum in powers of $t$. They are
\begin{align}
\label{IRpowerCounting}
&\mbox{$P_1$-collinear:} && k_{\rm c1} \sim \big(t^0, t, \sqrt{t} \big)  \, , \\
&\mbox{$P_2$-collinear:} && k_{\rm c2} \sim \big(t, t^0, \sqrt{t} \big)  \, , \\
&\mbox{ultrasoft:} &&  k_{\rm us} \sim \big(t, t, t \big)   \, .
\end{align}
These are the leading IR singular regions. Folding in the scaling of the integration measure \re{Dmeasure} along with the behavior of scalar 
propagators, one finds that $T_{11}$ has the following IR scalings 
\begin{align}
\label{T11scalingEq}
T^{\rm c1}_{11} \sim O (t^{-\ep})
\, , \qquad
T^{\rm c2}_{11} \sim O (t^{-\ep})
\, , \qquad
T^{\rm us}_{11} \sim O (t^{-2\ep})
\, , 
\end{align}
and asymptotes to
\begin{align}
\label{1LregionsC1}
T^{\rm c1}_{11} 
&
= -  \int_k D (k) D (k - P_1) D_{\rm eik} (k + P_2)
\, , \\
\label{1LregionsC2}
T^{\rm c2}_{11} 
&
= - \int_k D (k) D_{\rm eik} (k - P_1) D(k + P_2)
\, , \\
\label{1LregionsUS}
T^{\rm us}_{11} 
&
= - \int_k D (k) D_{\rm us} (k - P_1) D_{\rm us} (k + P_2)
\, .
\end{align}
Here, we introduced the eikonal and ultrasoft approximations for scalar propagators
\begin{align}
D_{\rm eik} (k_i \pm P_j) \equiv [\pm 2 k_i \cdot p_j ]^{-1} 
\, , \qquad
\label{USpropagator}
D_{\rm us} (k_i \pm P_j) \equiv [ P_j^2 \pm 2 k_i \cdot p_j ]^{-1} 
\, . 
\end{align}
To match the original integral \re{T11} to its IR decomposition, we have to account for an $t$-independent remainder. This corresponds to all 
components of the loop momentum being of order $O (t^0)$,
\begin{align}
&\mbox{hard:} && k_{\rm h} = \big(t^0, t^0, t^0 \big)  \, .
\end{align}
This defines the hard region, 
\begin{align}
T^{\rm h}_{11} 
= - \int_k D (k) D (k - P_1) D (k + P_2) |_{t = 0}
\end{align}
with the external virtuality exactly set to zero. In other words, its integrand is identical to the massless case. If we now allow for the (components 
of the) loop momentum to cover the entire real axis, we get the representation advocated within the Method of Regions \cite{Beneke:1997zp},
\begin{align}
T_{11} = T^{\rm h}_{11} + t^{-\ep} \big( T^{\rm c1}_{11} + T^{\rm c2}_{11} \big) + t^{-2 \ep} T^{\rm us}_{11}
\, . 
\end{align}
Here for presentation clarity, we explicitly pulled out the IR scalings \re{T11scalingEq} out of the integral definitions \re{1LregionsC1} -- \re{1LregionsUS}. 
Now, it becomes transparent why we insisted on keeping the dimension of space-time away from four. The above-approximated integrals
$T^{\rm c1}_{11}$, $T^{\rm c2}_{11}$ and $T^{\rm us}_{11}$ are divergent in the ultraviolet, while $m$ takes on the role of an IR `cutoff'. 
$T^{\rm h}_{11}$  is infrared divergent, on the other hand, and is regularized dimensionally. Explicit expressions for these integrals are well known
\cite{Becher:2014oda,Belitsky:2024yag} and are quoted for the reader's convenience in Appendix \ref{Appendix1L}. Summing up contributions of 
these regions, we observe the cancellation of the $\ep$-dependence and confirm Eq.\ \re{ExactOffSud} with just the leading $g^2$-terms kept in Eq.\ \re{ADs}.

It is important to stress that the Method of Regions has the status of experimental mathematics. However, up to now, there are no known 
examples where it is found to fail. We refer the interested reader to Refs.\ \cite{Semenova:2018cwy,Smirnov:2021dkb} for discussions of possible 
ways to prove this strategy.

This one-loop analysis suggests a multiplicatively factorized form for the near mass-shell form factor in terms of incoherent momentum 
components: hard, collinear, and ultrasoft
\begin{align}
\mathcal{F}_2 
= 
\left(1 - 2 g^2 T^{\rm h}_{11} \right) 
\left(1 - 2 g^2 t^{-\ep} T^{\rm c1}_{11} \right) \left(1 - 2 g^2 t^{-\ep} T^{\rm c2}_{11} \right) 
\left(1 - 2 g^2 t^{-2\ep} T^{\rm us}_{11} \right)
+
O (g^4)
\, ,
\end{align}
with each factor corresponding to its own momentum region and reflecting Taylor expansion of the integral $T_{11}$ in the vicinity of the 
corresponding pinch surfaces. Does this persist at higher loops? This is what we are set to answer in the rest of this paper.

Analyses of higher loop integrands with proper implementation of IR power counting \re{IRpowerCounting} is very challenging since it 
depends on the routing of loop momenta. So, an exhaustive enumeration of all leading regions is next to impossible. To circumvent this 
problem, the Feynman parameter integral representation is a far better choice \cite{Smirnov:1999bza}, and moreover, it can be recast in 
an entirely geometric way by associating a {\sl Newton polytope} to each integrand as a convex hull of its vertices \cite{Pak:2010pt}. We 
refrain from an in-depth review of this formalism since we recently did just that in Refs.\ \cite{Belitsky:2023ssv,Belitsky:2024yag}, but we 
will shed light on it from the perspective of Ref.\ \cite{Salvatori:2024nva}.

Since the two-loop integrals $T_{21}$ and $T_{22}$ were recently addressed in Ref.\ \cite{Belitsky:2024yag}, with results quoted in 
Appendix \ref{Appendix2L} to the required order $O (\ep^2)$, we jump straight to three loops.

\section{Solving integrals with Mellin-Barnes methods}

To reveal regions for a given Feynman integral, we apply a systematic algorithm \cite{Smirnov:1999bza,Pak:2010pt,Jantzen:2012mw} implemented 
in the computer code {\tt asy} \cite{Pak:2010pt,Jantzen:2012mw} that unambiguously determines lower facets of a Newton polytope connected 
with two Symanzik polynomials of the Feynman parametric representation. In fact, the expansion by regions can be applied with {\tt asy} to any 
parametric integral over\footnote{For other domains, one should first map it to $R_+^N$ and then proceed with {\tt asy}. Examples of its application 
to integrals, which are not Feynman integrals, can be found in \cite{Belitsky:2021huz} and \cite{Smirnov:2024pbj}.} $R_+^N$ with the integrand given 
by the product of polynomials raised to powers depending linearly on the regularization parameter $\varepsilon$. 

In the case of Feynman integrals, it proves more convenient to apply the code {\tt asy} as a part of the {\tt FIESTA5} distribution package 
\cite{Smirnov:2013eza,Smirnov:2021rhf} by executing the command {\tt SDExpandAsy} because within it all basic properties of Feynman 
integrals, including the folklore Cheng-Wu theorem\footnote{It manifests the GL(1) redundancy of the Feynman parameter integration measure.}
\cite{Cheng:1987ga}, are already taken into account. At this point, information about propagators, indices, and the desired orders of the 
$\varepsilon$-expansion and the expansion in the parameter $t$ is chosen.

In our previous paper~\cite{Belitsky:2023ssv}, we evaluated the asymptotic behavior of the three-loop integrals $T_{3n}$, shown in Fig.\ 
\ref{123IntegralsFig} for $n=1,2,\ldots,8$, at leading order in the limit $t \to 0$ as a terminating series in powers of $\log t$, with the largest 
power being $\log^6 t$. We relied mostly on the method of canonical differential equations \cite{Kotikov:1990kg,Gehrmann:1999as,Henn:2013pwa} 
with partial checks from the Method of Regions. Presently, our goal is to evaluate {\sl individual} contributions from all the regions in the expansion 
of the integrals $T_{3n}$. We can slightly simplify the task at hand by lumping up two integrals together, i.e., $T_{35}$ and $T_{36}$, since they have 
identical propagator structures, possess the same coefficients in the Sudakov form factor, and differ only in the presence of irreducible numerators. 
Thus, we evaluate the asymptotics of the combined integral $T_{3,56}=T_{35}+T_{36}$ from the same family with the following propagators 
and irreducible numerators
\begin{align}
\label{Props}
&
\{-k_1^2, -k_2^2, -k_3^2, -(k_3 - P_1)^2, -(k_2 + P_2)^2, -(k_1 + k_2 + k_3 - P_1)^2, 
-(k_1 + k_2 + k_3 + P_2)^2, 
\nonumber\\ &
-(k_2 + k_3 - P_1)^2, -(k_1 + k_2 + P_2)^2, -(k_3 - P_1 - P_2)^2, -(k_1 + k_2)^2, -(k_1 + k_3)^2 
\}
\, .
\end{align}      
Here, $k_i$ ($i = 1,2,3$) are the loop momenta, and the minus signs reflect our choice of the Euclidean kinematics. The $T_{3,56}$ is 
determined by the sum of its two Feynman integrals $F(1,1,\ldots,1,-1,0,0) +F(1,1,\ldots,1,0,-1,0)$.
 
To reveal all regions contributing at leading order in the limit $t\to 0$, we run {\tt SDExpandAsy} for the integrals $T_{3n}$, 
$i=1,2,3,4,56,7,8$. For example, in the $n=56$ case, we obtain 41 contributions corresponding to the regions with region vectors
\begin{align}
\bit{r}
=
\big\{
&
\{ 0, 0, 0, 0, 0, 0, 0, 0, 0, 0\}, 
\{0, 0, 1, 0, 0, 0, 1, 0, 0, 0 \}, 
\{ 0, 0, 1, 1, 1, 1, 2, 1, 1, 0\}, 
\nonumber\\
&
\{ 2, 0, 1, 1, 1, 2, 2, 1, 2, 0\}, 
\{ 1, 0, 1, 0, 1, 1, 2, 0, 1, 0\}, 
\{ 0, 1, 1, 0, 0, 1, 1, 1, 0, 0\}, 
\nonumber\\
&
\{1, 0, 1, 1, 1, 1, 2, 1, 2, 0 \}, 
\{1, 0, 1, 1, 0, 1, 1, 1, 1, 0 \}, 
\{ 1, 0, 1, 0, 0, 1, 1, 0, 1, 0\}, 
\nonumber\\
&
\{0, 0, 1, 1, 0, 1, 1, 1, 0, 0 \}, 
\{ 0, 0, 0, 0, 1, 0, 1, 0, 1, 0\}, 
\{ 0, 0, 0, 1, 1, 1, 1, 1, 1, 0\}, 
\nonumber\\
&
\{ 1, 0, 0, 0, 0, 1, 1, 1, 1, 0\}, 
\{0, 0, 1, 0, 1, 0, 2, 0, 1, 0 \}, 
\{ 2, 0, 1, 0, 1, 2, 1, 0, 1, 0\}, 
\nonumber\\
&
\{ 1, 1, 1, 0, 1, 1, 1, 1, 0, 0\}, 
\{2, 0, 1, 0, 1, 2, 2, 0, 2, 0 \}, 
\{ 1, 0, 0, 0, 1, 1, 1, 0, 1, 0\}, 
\nonumber\\
&
\{ 2, 1, 1, 0, 1, 2, 2, 1, 2, 0\}, 
\{1, 0, 1, 0, 1, 1, 1, 0, 0, 0 \}, 
\{ 1, 0, 0, 1, 1, 2, 0, 1, 0, 0\}, 
\nonumber\\
&
\{1, 1, 1, 0, 0, 2, 1, 1, 1, 0 \}, 
\{ 1, 0, 1, 1, 1, 2, 1, 1, 0, 0\}, 
\{ 1, 0, 0, 1, 1, 2, 1, 1, 1, 0\}, 
\nonumber\\
&
\{1, 1, 0, 0, 1, 1, 1, 1, 1, 0 \}, 
\{1, 1, 1, 1, 0, 2, 1, 2, 1, 0 \}, 
\{ 0, 0, 0, 0, 0, 1, 1, 1, 0, 0\}, 
\nonumber\\
&
\{ 1, 0, 1, 1, 1, 2, 2, 1, 1, 0\}, 
\{ 1, 1, 0, 1, 1, 2, 1, 2, 1, 0\}, 
\{ 1, 1, 0, 1, 1, 2, 0, 2, 0, 0\}, 
\nonumber\\
&
\{0, 1, 1, 1, 0, 2, 1, 2, 0, 0 \}, 
\{ 1, 1, 1, 1, 1, 2, 1, 2, 0, 0\}, 
\{2, 0, 1, 1, 1, 2, 1, 1, 1, 0 \}, 
\nonumber\\
&
\{ 1, 0, 0, 0, 0, 1, 0, 1, 0, 0\}, 
\{ 0, 1, 0, 0, 0, 1, 0, 1, 0, 0\}, 
\{1, 0, 0, 0, 1, 1, 0, 0, 0, 0 \}, 
\nonumber\\
&
\{ 1, 1, 1, 0, 1, 1, 2, 1, 1, 0\}, 
\{ 0, 0, 0, 1, 0, 1, 0, 1, 0, 0\}, 
\{ 1, 1, 0, 0, 1, 1, 0, 1, 0, 0\}, 
\nonumber\\
&
\{ 0, 1, 0, 1, 0, 2, 0, 2, 0, 0\}, 
\{ 2, 1, 1, 0, 1, 2, 1, 1, 1, 0\}
\big\}
\, .
\end{align}
In fact, their number remains intact at any order of the small $t$-expansion for this integral, i.e., they manifest themselves starting 
from the leading order, $t^0$, and running to $t^{- 6 \ep}$.The corresponding scalings associated with the above region vectors are
\begin{align}
\Big\{
&  
t^0, t^{-3 \varepsilon}, t^{-5 \varepsilon}, t^{-3 \varepsilon}, t^{-5 \varepsilon}, t^{-6 \varepsilon}, 
t^{-4 \varepsilon},  t^-\varepsilon, t^{-2 \varepsilon}, t^{-2 \varepsilon}, t^{-3 \varepsilon}, 
t^{-6 \varepsilon}, t^{-2 \varepsilon}, t^{-6 \varepsilon}, t^{-5 \varepsilon}, t^{-5 \varepsilon}, 
\nonumber\\
&
t^{-4 \varepsilon}, t^{-2 \varepsilon}, t^{-3 \varepsilon}, t^{-6 \varepsilon}, t^{-6 \varepsilon}, 
t^{-5 \varepsilon}, t^{-5 \varepsilon}, t^{-5 \varepsilon}, t^-\varepsilon, t^{-4 \varepsilon}, 
t^{-3 \varepsilon}, t^{-4 \varepsilon}, t^{-4 \varepsilon}, t^{-5 \varepsilon}, t^{-5 \varepsilon}, 
\nonumber\\
&
t^{-4 \varepsilon}, t^{-4 \varepsilon}, t^{-3 \varepsilon}, t^{-3 \varepsilon}, t^{-3 \varepsilon}, 
t^{-4 \varepsilon}, t^{-3 \varepsilon}, t^{-2 \varepsilon}, t^{-6 \varepsilon}, t^{-4 \varepsilon} 
\Big\} 
\, .
\end{align}
In this manner, we find Feynman parametric integrals that enter as their accompanying coefficients. It is our goal to calculate them
next as a Laurent series in the $\ep$-parameter. As usual, the use of the Cheng-Wu theorem is instrumental for the simplification of
required calculations. 

We postpone the discussion of the $t^0$-contribution till later and focus on the rest. We are left with the scalings $t^{-j\varepsilon}$ 
where $j=1,2,\ldots,6$. Our first approach to evaluate these parametric integrals in the $\ep$-expansion is the Mellin-Barnes (MB) 
method (see, e.g., Ref. \cite{Smirnov:2012gma}), which is based on the use of the simple formula 
\begin{align}
\frac{1}{(A+B)^{\lambda}} = \frac{1}{\Gamma(\lambda)}
\int_{\cal C} \frac{d z}{2 \pi i} 
\frac{B^z}{A^{\lambda+z}} \Gamma(\lambda+z)\Gamma(-z) \; .
\label{MB} 
\end{align}
It allows one to partition a complicated polynomial in Feynman parameters in terms of its two `simpler' components $A$ and $B$. In this 
equation, the contour ${\cal C}$ runs long the imaginary axis $-i \infty$ to $+i \infty$ of the complex $z$-plane such that all poles of 
$\Gamma(\ldots+z)$ are to its left while the ones of $\Gamma(\ldots-z)$ are to its right. Repeated use of the above formula allows one
to perform all parametric integrals in terms of Euler gamma functions, yielding a multifold MB integral. Of course, one attempts to arrive 
at as simple a final representation as possible with the fewest number of complex integrations. In most cases, we made this step with an
extensive use of the code {\tt MBcreate.m} \cite{Belitsky:2022gba}. However, sometimes it helped to derive MB representations by hand, 
especially in circumstances when one of the indices in a given integral was $-1$ since it inevitably led to very cumbersome numerators 
in contributing regions and visual lumping of terms was far more advantageous than the code could offer.

Having determined MB representations for all parametric integrals, the next step was to resolve singularities in $\varepsilon$ and construct
corresponding Laurent series. The goal here is to represent a given integral as a linear combination of MB integrals whose $\varepsilon$-expansions 
can be performed {\sl under} the integral sign. There are two public codes {\tt MB.m} and {\tt MBresolve.m} \cite{Czakon:2005rk,Smirnov:2009up}, 
where this algorithm is implemented automatically. Both of them are based on integration strategies developed in 
\cite{Tausk:1999vh,Smirnov:1999gc}. We relied on {\tt MBresolve.m} \cite{Smirnov:2009up}.

The next step is to evaluate MB integrals emerging as the coefficients in the Laurent expansion in $\varepsilon$. Here the command 
{\tt DoAllBarnes} from Kosower's\footnote{All required MB tools can be downloaded from {\tt https://gitlab.com/feynmanintegrals/MB}.} 
{\tt barnesroutines.m} automatically applies the first and the second Barnes lemmas (and their corollaries) whenever possible and thereby 
performs some integrations in terms of the Euler gamma functions. 

If, after all these possibilities were exhausted, some MB integrals are still left, one can turn to high-accuracy numerical evaluations and 
subsequent application of the {\tt PSLQ} algorithm~\cite{PSLQ:1999} to obtain analytic results in terms of a given basis of transcendental 
numbers. Here, we relied on the knowledge of uniform transcendental weights of all coefficients in the $\varepsilon$-expansion: the 
coefficients of $\varepsilon^{- 6}$ are just rational numbers, the terms with $\varepsilon^{-5}$ are absent (since we use the standard 
normalization of Feynman integral by multiplying the result by $e^{\gamma_{E} \varepsilon}$ per loop), the terms with $\varepsilon^{-4}$, 
$\varepsilon^{-3}$, $\varepsilon^{-2}$, are proportional to $\zeta_2$, $\zeta_3$, $\zeta_4$, respectively, while $\varepsilon^{-1}$, and 
$\varepsilon^{0}$ are accompanied by a linear combination of $\zeta_3 \zeta_2$ and $\zeta_5$, and, respectively, $\zeta_3^2$ and $\zeta_6$ 
with rational coefficients. For one-fold MB integrals, one can achieve an arbitrarily high precision, so that an analytic result is guaranteed. 
Our experience shows that, for two-fold MB integrals, the precision of 20 decimal places is sufficient to arrive at an analytic result as well. 
Moreover, we have found that in this fashion, even three-fold integrals can be successfully tackled as well, since sufficiently high precision 
can indeed be reached. However, above that, this strategy is hopeless. So we had to turn to other means, as we describe next. Those 
cases were analyzed using the method introduced in \cite{Salvatori:2024nva} as well as the method of canonical differential equations. 

\section[Solving integrals with tropical geometry]{Solving integrals with tropical geometry\footnote{This section was made possible largely 
due to a collaboration with Giulio Salvatori. His method, introduced in Ref.\ \cite{Salvatori:2024nva}, was indispensable for the successful 
calculation of contributions of some of the most complicated regions at three-loop order.}}

Let us provide a brief introduction to the formalism of Ref.\ \cite{Salvatori:2024nva}, which was instrumental in finding the $\ep$-expansion of 
contributions of some relevant regions. The region integrals are examples of the so-called {\sl Euler integrals} that admit the following generic form
\begin{align}
I({\bf c}; {\bf{s}}) 
= 
\int_{\mathbb{R}^{n}_{\ge 0}} \frac{d\alpha}{\alpha} \prod_{i=1}^m 
P_j(\alpha, s)^{c_j}, \quad 
\frac{d\alpha}{\alpha} \equiv \frac{d\alpha_1}{\alpha_1}\dots \frac{d\alpha_n}{\alpha_n}
\, ,
\label{eq:euler}
\end{align}
where $P_j(\alpha, s)$ are polynomials in the integration variables with coefficients collectively denoted by $\bf s$,
\begin{align}
P_j(\alpha, s) 
= 
\sum_{{\bf m} \in \mathbb{Z}^n} s_{\bf m} \alpha^{\bf m}, \quad \alpha^{\bf m}\coloneqq \alpha_1^{{\bf m}_1} \dots \alpha_n^{{\bf m}_n}.
\end{align}
In the present case, all exponents $c_j$ depend linearly on the common parameter $\ep$, i.e.,  the dimensional regulator, $c_j = c_{j,1} \ep + c_{j,0}$.

Since all of the expansion coefficients ${\bf s}$ are non-negative in the current application to the Sudakov integrand, singularities arise solely from
the boundaries of the integration regions. There are none in the bulk! The Laurent expansion of $I(\ep,{\bf s})$ in $\ep$ can then be obtained with 
the so-called {\sl `subtraction' formula} introduced in Ref.\ \cite{Salvatori:2024nva}. This allows us to express $I(\ep,{\bf s})$ as a linear combination of 
{\sl locally finite} integrals, i.e., they can be expanded in $\ep$ directly {\sl under} the integral sign and then integrated term by term\footnote{Local 
finiteness is a more stringent condition than just finiteness, which is merely the statement of the absence of poles in $\ep$ in the integrated result.}.

In order to apply the aforementioned formula, conditions of the corresponding Theorem~6 from Ref.\ \cite{Salvatori:2024nva} must be met. These can 
be formulated as geometrical properties satisfied by the Newton polytope of Symanzik polynomials determining the integrand,
\begin{align}
\mathcal{P} \coloneqq \mathrm{Newt\ } \prod_{j=1}^{\bf m} P_j
\, .
\label{eq:newton}
\end{align}
The singularity structure of Feynman integrals is related to the facet, rather than the aforementioned vertex, description of the associated 
Newton polytope \cite{Arkani-Hamed:2022cqe,Hillman:2023vas}. It becomes indispensable in the current consideration, and we refer to 
\cite{Arkani-Hamed:2022cqe,Salvatori:2024nva} for a precise definition of the corresponding cutout conditions. Let us consider a collection of 
vectors $\rho \in \mathbb{Z}^n$  that are outward-pointing normals to the facets, i.e., co-dimension one faces, of $\mathcal{P}$. These define 
the boundaries of domains of linearity. We say that two rays $\rho$ and $\rho'$ are {\sl compatible} if the corresponding facets of $\mathcal{P}$ 
intersect in a lower-dimensional face of the polytope. For each vector $\rho$, we consider a rescaling of the integration variables 
$\alpha_i \to \alpha_i \lambda^{-\rho_i}, i =1, \dots, n$, under which the integrand of \eqref{eq:euler} 
transforms as
\begin{align}
\prod_{j=1}^m P_j(\alpha_i \lambda^{-\rho_i}) = \lambda^{-\mathrm{Trop}(\rho)} \times \mathcal{J}(\lambda)
\, ,
\end{align}
where $\mathcal{J}(\lambda)$ is a holomorphic function of $\lambda$. The exponent $\mathrm{Trop}(\rho)$ controls the divergence of the 
integral in the vicinity of a corner of the integration domain that was exposed by the rescaling. If $\mathrm{Trop}(\rho)=a + b\ep$ and $a>0$, 
the integrand has a power-like divergence, if $a=0$ it possesses a logarithmic divergence, and if $a<0$ the integrand is finite at that corner. 
From now on, we will solely focus on the divergent rays and their corresponding facets.

The conditions of Theorem 6 require that
\begin{enumerate}
\item $a=0$ for all divergent rays (log-divergences only).
\item No divergent ray can be expressed as a linear combination of the divergent rays that are compatible with it, i.e., all cones that they span are
simplicial.
\end{enumerate}
Assuming these are satisfied, Theorem 6 of \cite{Salvatori:2024nva} states that the following `subtraction' formula holds
\begin{align}
I({\bf c}; {\bf s}) = \sum_{F \in \partial P} \mathrm{Vol}(F) I_F^{\rm ren} 
\, .
\label{eq:subformula}
\end{align}
Here, the sum runs over divergent faces of $\mathcal{P}$ of all possible dimensions. If a face $F$ is the intersection of several facets $\rho$ it 
contributes with a ``volume"  \cite{Arkani-Hamed:2022cqe} (see also \cite{Hillman:2023vas})
\begin{align}
\mathrm{Vol}(F) = \prod_{\rho \in F} \frac{1}{\mathrm{Trop}(\rho)},
\end{align}
which is responsible for an overall pole in $\ep$, due to the fact that $\mathrm{Trop}(\rho)=\mathcal{O}(\ep)$ on a divergent ray. The volume factor 
multiplies an integral $I_F^{\rm ren}$, which is of the same dimensionality as the face $F$. The integrand of $I_F^{\rm ren}$ is constructed starting 
from the integrand of $I({\bf c}, {\bf s})$ and `restricting' it on the face $F$, i.e., keeping in each polynomial $P_j$ only the monomials that live on the 
face $F$. The resulting integrand can then be made finite by subtracting from it contributions coming from faces of the polytope that are compatible 
with $F$. This procedure ensures that the integrals $I_F^{\rm ren}$ are locally finite. Therefore, we can evaluate all contributions appearing in the 
right-hand side of the `subtraction' formula by simply expanding all {\sl integrands} in $\ep$ and evaluating emerging integrals term-by-term.

If one of the two conditions of the theorem is not satisfied, \cite{Salvatori:2024nva} provides a lifeline. For divergences stronger than logarithmic, 
one can apply the Nilsson-Passare analytical continuation\footnote{The Nilsson-Passare procedure is based on a repeated integration-by-parts along
facets' normals and is similar in spirit to the original integration-by-parts of 't Hooft and Veltman, which allows one to construct a meromorphic 
continuation of momentum integrals in space-time dimension \cite{tHooft:1972tcz}.} \cite{Berkesch,Nilsson} on the rays that break the first criterion, 
resulting in an expansion of the original integrand into a linear combination of integrands that do satisfy the criterium. This approach also helps if 
the second condition is not fulfilled. Alternatively, if one encounters a non-simplicial cone, one needs to triangulate it into simplicial ones. However, 
the fact that we generally deal with polytopes that are not permutahedra, one needs to perform this non-unique decomposition on a case-by-case basis. 
The author of \cite{Salvatori:2024nva} warns us that this is, in general, an expensive procedure, and indeed, we confirm this in our calculation, as we 
will explain shortly.

Having reviewed the formalism of \cite{Salvatori:2024nva}, we now explain how we have applied it to the problem at hand. We have constructed 
Newton polytopes for each region integral using the code {\tt polymake} of Ref.\ \cite{Gawrilow:2000qhs}. For the integrals satisfying the requirements 
alluded to above, we then applied the `subtraction' formula \eqref{eq:subformula} \cite{Salvatori:2024nva}.

\begin{figure}
\centering
\begin{subfigure}[c]{0.45\textwidth}
\centering
\vspace{1cm}
$
\left(
\begin{array}{c@{\hspace{8pt}}c@{\hspace{4pt}}c@{\hspace{4pt}}c@{\hspace{4pt}}}
\left(
\begin{array}{r} 0 \\ 0 \\ 1 \\ 1 \\ 0 \\ 0 \\ 0 \\ 0 
\end{array}
\right)  
\left(
\begin{array}{r} 
0 \\ -1 \\ 0 \\ -1 \\ 0 \\ -1 \\ -1 \\ -1 
\end{array}
\right) 
\left(
\begin{array}{r} 
0 \\ 0 \\ 1 \\ 0 \\ 1 \\ 0 \\ 0 \\ 0 
\end{array}
\right) 
\left(
\begin{array}{r} 
0 \\ 1 \\ 1 \\ 1 \\ 0 \\ 0 \\ 0 \\ 1 
\end{array}
\right) 
\end{array} 
\right)
$
\label{fig:matrix}
\end{subfigure}
\begin{subfigure}[c]{0.45\textwidth}
\centering
\includegraphics[width=0.9\linewidth]{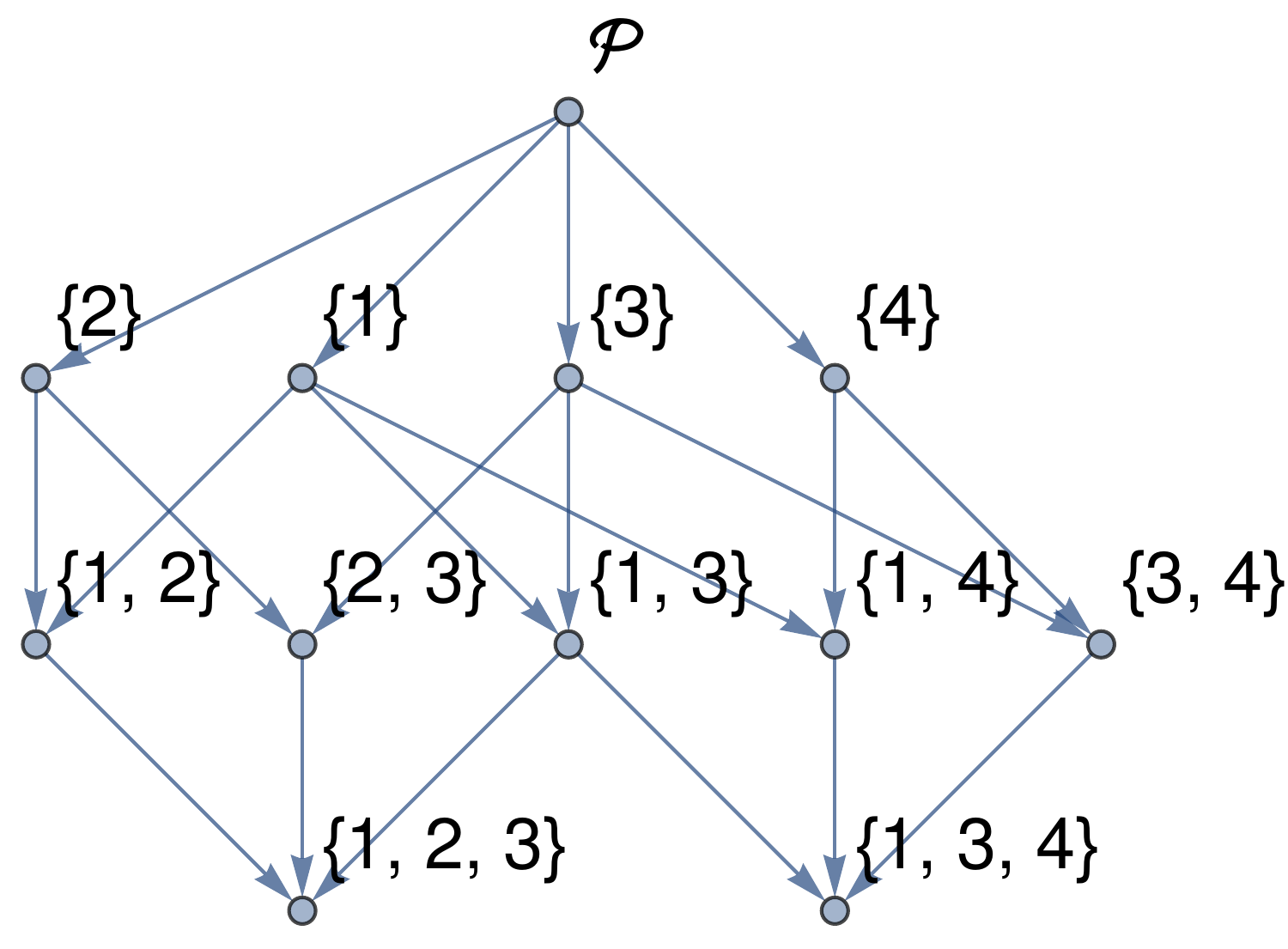}
\label{fig:example-image}
\end{subfigure}
\hfill
\caption{Divergent rays of a region contributing to diagram $T_{33}$, listed as columns of a matrix (left panel). The lattice of divergent facets of 
the corresponding Newton polytope (right panel). Each node represents a face of the polytope where the facets with the listed normals meet.}
\label{fig:raysAndFaces}
\end{figure}

As an example, for one of the regions contributing to the graph $T_{32}$, namely with the region vector $\bit{r} = \{0, 0, 0, 0, 1, 1, 1, 0, 1, 0\}$, we 
find the divergent rays shown in the left panel of Fig.~\ref{fig:raysAndFaces}. These are the first integer vectors along the positive direction of 
the outward-pointing normal to a facet of the Newton polytope $\mathcal{P}$ defined in Eq.\ \re{eq:newton}. We listed only facets that are 
responsible for a pole in $\ep$, i.e., such that $\mathrm{Trop}(\mathcal{I})[\rho]=\mathcal{O}(\ep)$. The subset of $\partial\mathcal {P}$ inducing 
higher order poles in $\ep$ is shown in its combinatorial structure in the right panel of Fig.~\ref{fig:raysAndFaces}. These are 
dual to the cones formed by taking a positive combination of compatible rays. Each node in the partially ordered set corresponds 
to a face $F$ of $\mathcal{P}$ of co-dimension equal to the number of compatible rays. In particular, at the lower level of the 
partially ordered set, we find the cones responsible for the leading pole in $\ep$. They are computed by means of the formula 
\eqref{eq:subformula} from \cite{Salvatori:2024nva} in terms of convergent integrals of the lowest dimension. Conversely, the top level of the diagram is a single integral of the 
highest dimension. Its integrand is the result of the renormalization map, described in Ref.\ \cite{Salvatori:2024nva}, applied to 
the integrand \eqref{eq:euler}. This is the only term in \eqref{eq:subformula} that does not produce a pole in $\ep$.

As a next step, we evaluated emerging locally finite integrals from the previous step by expanding their integrands in $\ep$ and integrating the 
series term-by-term using {\tt HyperInt} of \cite{Panzer:2014gra}. The algorithm devised in \cite{Panzer:2014gra} integrates one variable at a time, 
expressing the result in terms of Goncharov polylogarithms. This works only for {\sl linearly reducible} integrals, those for which there is at least 
one order of the integration variables where at each step the partial integral is a polylogarithmic function with rational prefactors such that both 
the letters and the poles of the prefactor are linear in the next integration variable. An appealing feature of this code is that it provides a way to 
safely ignore spurious divergent contributions appearing in intermediate integration steps with the use of the shuffle regularization. This means 
that if we know that an integral $\int\left(\mathcal{I}_1-\mathcal{I}_2\right)$ is convergent, even though the individual integrals $\int \mathcal{I}_i$ are divergent, 
{\tt HyperInt} is able to integrate each integrand separately while systematically dropping divergent terms. The latter are guaranteed to cancel when 
adding the two up together. The only restriction to this procedure is that both integrals must be calculated by integrating variables in the {\sl same} 
order. Accordingly, in order to evaluate an integral of a locally finite integrand, written as a combination of Euler integrands in the same integration
domain, we listed all the admissible integration orders for each of the individual integrands and then kept only those that are common.

In principle, \verb|HyperInt| should be able to integrate locally finite integrands using any common order. However, different orders required very 
different CPU times for the integrals to be carried over, with some simply never terminating. As a practical solution to this issue, we merely attempted 
various orders simultaneously until we found one that terminated successfully.

With this strategy, we were able to evaluate all the contributions of regions for diagrams $T_{3,1}$, $T_{3,2}$, $T_{3,3}$, $T_{3,4}$, and $T_{3,7}$. Among 
these, the regions $T_{31}^{\rm h\text{-}h\text{-}us}$, $T_{31}^{\rm c\text{-}c\text{-}us}$,  $T_{32}^{\rm h\text{-}h\text{-}h}$ and  $T_{32}^{\rm h\text{-}h\text{-}c}$ 
proved to be untractable by the Mellin-Barnes approach described in the previous section.

On the other hand, not all of the region integrals met the above criteria. For example, for the diagram $T_{3,56}$ the region with vector
$\bit{r} = \{1, 1, 1, 0, 1, 1, 2, 1, 1, 0\}$ yields an integral whose divergent rays are depicted in Fig.~\ref{fig:nonplanar}.
\begin{figure}
\centering
\[
\left(
\begin{array}{c@{\hspace{8pt}}c@{\hspace{8pt}}c@{\hspace{8pt}}c@{\hspace{8pt}}c@{\hspace{8pt}}c@{\hspace{8pt}}c@{\hspace{8pt}}c@{\hspace{8pt}}c}
\left(\begin{array}{r}
1 \\ 1 \\ 0 \\ 0 \\ 0 \\ 1 \\ 0 \\ 1
\end{array}\right) &
\left(\begin{array}{r}
0 \\ 1 \\ 0 \\ 0 \\ 0 \\ 0 \\ 0 \\ 1
\end{array}\right) &
\left(\begin{array}{r}
0 \\ 0 \\ 1 \\ 0 \\ 0 \\ 0 \\ 1 \\ 0
\end{array}\right) &
\left(\begin{array}{r}
0 \\ 1 \\ 1 \\ 0 \\ 1 \\ 0 \\ 1 \\ 0
\end{array}\right) &
\left(\begin{array}{r}
0 \\ 0 \\ 0 \\ -1 \\ 0 \\ -1 \\ 0 \\ -1
\end{array}\right) &
\left(\begin{array}{r}
0 \\ 0 \\ -1 \\ -1 \\ 0 \\ 0 \\ 0 \\ 0
\end{array}\right) &
\left(\begin{array}{r}
0 \\ 1 \\ 1 \\ 1 \\ 1 \\ 0 \\ 1 \\ 1
\end{array}\right) &
\left(\begin{array}{r}
0 \\ -1 \\ -1 \\ -1 \\ 0 \\ -1 \\ 0 \\ -1
\end{array}\right) &
\left(\begin{array}{r}
-1 \\ -1 \\ -1 \\ -1 \\ -1 \\ -1 \\ 0 \\ -1
\end{array}\right)
\end{array}
\right)
\]
\caption{The normals to the Newton polytope for a region of diagram $T_{3,56}$, listed as columns of a matrix. }
\label{fig:nonplanar}
\end{figure}
The fourth divergent ray is compatible with all the others, and it belongs to their linear span, breaking the second criterion.
Applying once a Nilsson-Passare analytical continuation \cite{Berkesch,Nilsson} along the fourth vector, we reduce the original 
integral to a combination of integrals satisfying the required criteria. In this manner, we can therefore calculate them individually 
with the same formula as before. The caveat is that we had to reach a higher order in the $\ep$-expansion. Although we always 
find them to be linearly reducible, the sheer amount of integrands adds up quickly and becomes very time-consuming to be 
a practical strategy\footnote{To address this issue, we consider it highly desirable to develop a proper parallelization of core 
functionalities of {\tt HyperInt}.}. This issue was particularly severe for diagrams $T_{3,56}$ and $T_{3,8}$, for which multiple 
regions broke the requirements necessary for the `subtraction' formula. Luckily, the contributions of regions for $T_{3,8}$ proved 
tractable by the Mellin-Barnes approach, and the contributions of regions for $T_{3,56}$ were calculated with differential equations,
so that every region was covered by at least one of the three methods.
 
\section{Solving integrals with differential equations}

We now return to the remaining, hard regions with $t^0$-behavior. These are obtained from ab initio integrals by setting $t=0$. Their integrands
do not simplify sufficiently for $t=0$ to warrant using their parametric representations directly for subsequent calculations. Instead, it is much more 
efficient to reduce the original massless on-shell momentum integrals to a set of master integrals by means of integration-by-parts relations. These 
master integrals were calculated before \cite{Gehrmann:2006wg,Heinrich:2007at,Baikov:2009bg,Heinrich:2009be} and are available up to 
transcendentality weight eight from Ref.\ \cite{Lee:2010ik}.
 
Finally, we needed to calculate contributions of individual regions for integrals that yielded multifold MB integrals that could not be tamed numerically 
with sufficient precision. All of them appeared in $T_{3,56}$. Here again, we have applied the method of the differential equations 
\cite{Kotikov:1990kg,Gehrmann:1999as,Henn:2013pwa}. In fact, this method was applied earlier to the three-loop massless vertex 
integrals~\cite{Pikelner:2021goo} at the so-called symmetrical point, i.e., $p_1^2 = p_2^2 = q^2 = -1$. By adopting this earlier analysis in 
\cite{Belitsky:2023ssv}, we evaluated our integrals $T_{3n}$ at general values of $Q^2$ and $m^2$, with $p_1^2 = p_2^2=-m^2$ and then 
deduced results for their $t \to 0$ asymptotics by expanding the thus-obtained analytical results.

Presently, we need to evaluate, however, the contributions of {\sl individual} regions for the leading order in the small-$t$ expansion of $T_{3,56}$.
As in the previous analysis, we start from differential equations at generic values of $t$ for the 77 primary master integrals. We introduce a new 
variable, i.e, $t = -(z-1)^2/z$, that rationalizes and, therefore, avoids potential square roots in resulting differential equations. Then turn to their 
canonical form \cite{Henn:2013pwa} using the available public codes like, e.g., {\tt CANONICA} \cite{Meyer:2017joq},
\begin{align}
\frac{\partial}{\partial t} \bit{f} = \varepsilon \bit{A} \cdot \bit{f}
\, .
\end{align} 
Here $\bit{f}$ is a vector of canonical basis elements and $\bit{A}$ is a corresponding matrix with elements being functions of $z$ but 
independent of $\varepsilon$. 
 
Since we only need the leading asymptotics of the 77 elements of the canonical basis in the limit $t\to 0$, these can be extracted from the matrix 
exponent $\exp\left(\bit{A}_0 \log t \right)\cdot \bit{b}$, where $\bit{A}_0 $ is the $1/t$-coefficient of $\bit{A}$. The rest of the terms are irrelevant for 
our purposes. Matrix elements of this matrix exponent are linear combinations of the powers $t^{-j \varepsilon}$ with $j=0,1,\ldots,6$.
Here $b_i = \{ \bit{b} \}_i$, $i=1,2,\ldots,77$, are the integration constants. They are uniformly-transcendental linear combinations of the form $b_{ij} 
\varepsilon^j$, where $j=-2,-1,\ldots 4$. Once these boundary conditions are fixed, we obtain the leading asymptotics of the elements of the 
canonical basis in the limit $t\to 0$. Integration-by-parts identities provide us with a relation of $T_{3,56}$ to the found elements of the canonical basis in 
the limit $t \to 0$. To fix $b_{i}$, it was sufficient to apply explicit analytic solutions for the master integrals at general $t$ obtained in~\cite{Pikelner:2021goo}
and then to expand corresponding expressions in the leading order of the limit $t\to 0$.

In this manner, we deduced the Laurent expansion of contributions to all relevant regions up to $O (\ep^0)$. They are listed in Appendix \ref{Appendix3L}. 
A more detailed presentation of three-loop results is given in the accompanying ancillary files attached with this submission.

\section{Factorization}

Now we are in a position to assemble our findings in the three-loop form factor. Contributing graphs, enumerated according to the nomenclature of
Fig.\ \ref{123IntegralsFig}, at each consecutive $\ell$-loop order, enter the Sudakov form factor with integer coefficients $\alpha_{\ell n}$. 
$\mathcal{F}_2^{(1)}$ was quoted in Eq.\ \re{F21}, so $\alpha_{11} = -2$. For the rest, we have
\begin{align}
\label{F22}
\mathcal{F}_2^{(2)} 
= \sum_{n = 1}^2 \alpha_{2n} T_{2n}
\, , \qquad
\mathcal{F}_2^{(3)} 
= \sum_{n = 1}^8 \alpha_{3n} T_{3n}
\, ,
\end{align}
with
\begin{align}
\alpha_{2n} = \{ 4 , 1 \}
\, , \qquad
\alpha_{3n} = \{ - 8, - 2, 4, 4, -4, -4, -4, 2 \}
\, .
\end{align}
Here, the two- and three-loop integrals are decomposed over contributing regions as
\begin{align}
T_{2n} 
&= 
T^{\rm h\text{-}h}_{2n} + t^{-\ep} T^{\rm h\text{-}c}_{2n} 
+ 
t^{-2 \ep} \left( T^{\rm c\text{-}c}_{2n} + T^{\rm h\text{-}us}_{2n} \right)
+ 
t^{-3 \ep} T^{\rm c\text{-}us}_{2n} + t^{-4 \ep} T^{\rm us\text{-}us}_{2n} 
\, , \\
T_{3n} 
&= 
T^{\rm h\text{-}h\text{-}h}_{3n} + t^{-\ep} T^{\rm h\text{-}h\text{-}c}_{3n} 
+ 
t^{-2 \ep} \left( T^{\rm h\text{-}c\text{-}c}_{3n} + T^{\rm h\text{-}h\text{-}us}_{3n} \right)
+ 
t^{-3 \ep} \left( T^{\rm c\text{-}c\text{-}c}_{3n} + T^{\rm h\text{-}c\text{-}us}_{3n} \right) 
\\
&\qquad\qquad\!
+ 
t^{-4 \ep} \left( T^{\rm c\text{-}c\text{-}us}_{3n} +  T^{\rm h\text{-}us\text{-}us}_{3n} \right)
+
t^{-5 \ep} T^{\rm c\text{-}us\text{-}us}_{3n} 
+
t^{-6 \ep} T^{\rm us\text{-}us\text{-}us}_{3n} 
\, , \nonumber
\end{align}
where we did not split collinear regions up further with respect to legs $P_1$ and $P_2$. We can immediately verify the correctness
of the total sum against Eq.\ \re{ExactOffSud}, which was obtained solely on the basis of determining master integrals through the
method of differential equations. Indeed, it confirms the exponentiation of the Sudakov form factor flawlessly.

Anticipating the hard-ultrasoft-collinear factorization, we write an ansatz for $\mathcal{F}_2$
\begin{align}
\label{MultiplicativeFactEq}
\mathcal{F}_2 = h (\ep) J \big( \ep, \sqrt{t} \big) S (\ep, t)
\, ,
\end{align}
where the hard $h$, jet $J$, and ultrasoft $S$ functions receive contributions solely from hard, collinear, and ultrasoft modes, respectively. Explicitly,
\begin{align}
\label{hEq}
h (\ep) 
&
= 
1 + g^2 \alpha_{11} T_{11}^{\rm h} 
+ 
g^4 \sum_{n = 1}^2 \alpha_{2n} T_{2n}^{\rm h\text{-}h} + g^6 \sum_{n = 1}^8 \alpha_{3n} T_{3n}^{\rm h\text{-}h\text{-}h} 
+ 
O (g^8)
\, , \\
\label{JEq}
J \big( \ep, \sqrt{t} \big)
&
= 
1 + g^2 \alpha_{11} t^{- \ep} T_{11}^{\rm c} 
+ 
g^4 t^{- 2\ep} \sum_{n = 1}^2 \alpha_{2n} T_{2n}^{\rm c\text{-}c} + g^6 t^{-3 \ep} \sum_{n = 1}^8 \alpha_{3n} T_{3n}^{\rm c\text{-}c\text{-}c} 
+ 
O (g^8)
\, , \\
\label{SEq}
S ( \ep, t )
&
= 
1 + g^2 \alpha_{11} t^{- 2 \ep}T_{11}^{\rm us} 
+ 
g^4 t^{- 4 \ep}\sum_{n = 1}^2 \alpha_{2n} T_{2n}^{\rm us\text{-}us} + g^6  t^{- 6 \ep} \sum_{n = 1}^8 \alpha_{3n} T_{3n}^{\rm us\text{-}us\text{-}us} 
+ 
O (g^8)
\, ,
\end{align}
where $T_{11}^{\rm c} \equiv T_{11}^{\rm c1} + T_{11}^{\rm c2}$ from Eqs.\ \re{1LregionsC1} and \re{1LregionsC2}.
This form implies that all mixed contributions, such as for instance, $T^{\rm h\text{-}c}$, should arise from products of individual regions 
from lower loop order. This statement can be confirmed or rebutted with our explicit three-loop results.

\subsection{Hard factorization}
\label{HardSect}

Let us begin this analysis with the mixed contributions involving at least one hard region. At two loops, it is not hard to check that
\begin{align}
\sum_{n = 1}^2 \alpha_{2n} T_{2n}^{\rm h\text{-}c} = 4 T_{11}^{\rm h} T_{11}^{\rm c} 
\, , \qquad
\sum_{n = 1}^2 \alpha_{2n} T_{2n}^{\rm h\text{-}us} = 4 T_{11}^{\rm h} T_{11}^{\rm us}
\, .
\end{align}
Analogously, we find at three loops
\begin{align}
&
\sum_{n = 1}^8 \alpha_{3n} T_{3n}^{\rm h\text{-}h\text{-}c} = - 2 T_{11}^{\rm c} \sum_{n=1}^2 \alpha_{2n}T_{2n}^{\rm h\text{-}h} 
\, , \qquad
\sum_{n = 1}^8 \alpha_{3n} T_{3n}^{\rm h\text{-}h\text{-}us} = - 2 T_{11}^{\rm us} \sum_{n=1}^2 \alpha_{2n}T_{2n}^{\rm h\text{-}h} 
\, , \\
&
\sum_{n = 1}^8 \alpha_{3n} T_{3n}^{\rm h\text{-}c\text{-}c} = - 2 T_{11}^{\rm h} \sum_{n=1}^2 \alpha_{2n}T_{2n}^{\rm c\text{-}c}
\, , \qquad
\sum_{n = 1}^8 \alpha_{3n} T_{3n}^{\rm h\text{-}us\text{-}us} = - 2 T_{11}^{\rm h} \sum_{n=1}^2 \alpha_{2n}T_{2n}^{\rm us\text{-}us} 
\, , \\
&
\label{hcUS}
\qquad\qquad\qquad\qquad\qquad
\sum_{n = 1}^8 \alpha_{3n} T_{3n}^{\rm h\text{-}c\text{-}us} = - 2 T_{11}^{\rm h} \sum_{n=1}^2 \alpha_{2n}T_{2n}^{\rm c\text{-}us} 
\, .
\end{align}
These relations prove the expected factorization of the hard region from the ones with ultrasoft-collinear modes up to three-loop order.

\subsection{Ultrasoft-collinear factorization: a two-loop obstruction}
\label{UScolSect}

We now turn to separating collinear from ultrasoft regions. A quick inspection shows, however, that it is broken starting from
two loops. Namely,
\begin{align}
\label{cUSbrokenEq}
\sum_{n = 1}^2 \alpha_{2n} T_{2n}^{\rm c\text{-}us} \neq 4 T_{11}^{\rm c} T_{11}^{\rm us} 
\, ,
\end{align}
and this implies that the alleged multiplicative factorization \re{MultiplicativeFactEq} is violated.

Let us zoom in on this `problematic' region. Since the double ladder and cross ladder yield, up to an overall numerical factor, 
identical expressions for this region, i.e., $4 T_{21}^{\rm c\text{-}us} = T_{22}^{\rm c\text{-}us}$, we can focus on just one of 
these integrals. We consider $T_{21}$ which reads
\begin{align}
\label{T21}
T_{21}
&
=
\int_{k_1, k_2} D(k_1) D(k_2) D(k_1 - P_1) D (k_1 + k_2 - P_1) D (k_1 + P_2) D (k_1 + k_2 + P_2)
\, .
\end{align}
In the ${k_2\text{-}k_1}$-ordering there are two identical contributions, i.e., ${\rm c1\text{-}us}$ and ${\rm c2\text{-}us}$, defining 
the ultrasoft-collinear region. Again, it suffices to discuss just one, say, ${\rm c1\text{-}us}$ determined by the vector 
$\bit{r}_{\rm c1\text{-}us} = \{0, 1, 1, 1, 1, 2\}$, which unambiguously specifies the IR scaling of the integrand's propagators.
Namely, according to IR counting rules \re{IRpowerCounting}, $D(k_1)$ and $D(k_2)$ behave as
\begin{align}
D(k_1) \sim t^{- 2}
\, , \qquad
D(k_2) \sim t^{- 1}
\, ,
\end{align}
and do not change their form since both terms in the light-cone decomposition $k_i^2 =  k_{i \perp}^2 - k_{i+} k_{i -}$ scale the same
way. The propagator $D (k_1 + k_2 + P_2) \sim t^0$ admits the eikonal form
\begin{align}
D (k_1 + k_2 + P_2) = D_{\rm eik} (k_2 + P_2)
\, ,
\end{align}
while $D (k_1 - P_1) \sim t^{- 1}$ and $D (k_2 + P_2) \sim t^{- 1}$ are ultrasoft in this region 
\begin{align}
\label{k1USprop}
D (k_1 - P_1) = D_{\rm us} (k_1 - P_1)
\, , \qquad
D (k_1 + P_2) = D_{\rm us} (k_1 + P_2)
\, .
\end{align}
Finally, $D (k_1 + k_2 - P_1) \sim t^{-1}$ possesses two additive contributions
\begin{align}
\label{Offender}
D (k_1 + k_2 - P_1) = [ (k_2 - p_1)^2 + 2 k_1 \cdot (k_2 - p_1) ]^{-1}
\, .
\end{align}
To achieve the desired factorization of the ultrasoft from the collinear region, we would like to drop the second term, which entangles 
the two loop momenta. However, this is not justified since both of them provide parametrically leading-order effects according 
to the IR counting rules. 

To unravel the effect of this intertwining, we calculate the ultrasoft loop first. To this end, we use the Schwinger-time representation
for the ultrasoft propagators \re{k1USprop} and the `offending' one \re{Offender}, where we temporarily introduce $K_2 = k_2 - p_1$,
\begin{align}
\mathcal{L}_1
&
=
\int_{k_1} 
D(k_1) D_{\rm us} (k_1 - P_1) D_{\rm us} (k_1 + P_2) D (k_1 + K_2)
\nonumber\\
&
=
- i \e^{\ep \gamma_{\rm E}}
\int_0^\infty d \alpha \e^{i \alpha K_2^2}
\int_0^\infty
d \beta_1 d \beta_2 \e^{- i t (\beta_1 + \beta_2)} 
\frac{\Gamma (1 - \ep)}{[- 2 \beta_2 p_2 \cdot (\alpha K_2 - \beta_1 p_1)]^{1 - \ep}}
\, ,
\end{align}
where we extracted the leading IR asymptotics in the denominator after the Fourier transform in $k_1$. Next, integrating over 
$\beta_2$ is straightforward as well as $\alpha$ after the rescaling $\beta_1 \to \alpha \beta_1$. Finally, changing the 
integration variable $\beta_1$ to $x$ via the map $\beta = \bar{x}/x$, we get
\begin{align}
\mathcal{L}_1
=
\e^{\ep \gamma_{\rm E}}
\frac{\Gamma (\ep) \Gamma (1 - \ep) \Gamma (1 + \ep)}{t^{2 \ep} [- K_2^2]} 
\int_0^1 d x \bar{x}^{\ep - 1} \left[ 1 + 2 t \frac{x}{\bar{x}} \frac{p_2 \cdot K_2}{[- K_2^2]} \right]^{\ep - 1}
\, .
\end{align}
We now see explicitly that the second term in the square bracket of the integrand is the obstruction for factorization. It cannot 
be ignored since, while its numerator is suppressed by $t$, it gets enhancement from the $P_1$-collinear kinematics of the 
denominator. It is easy to calculate the resulting $x$-integral yielding a closed-form geometric series. However, we will not do it 
since the integral form is more appropriate for the subsequent evaluation of the $k_2$-loop. It yields the following $\ep$-exact
expression
\begin{align}
T_{21}^{\rm c1\text{-}us}
=
\e^{2 \ep \gamma_{\rm E}} \frac{\Gamma^3 (\ep) \Gamma (- \ep) \Gamma^2 (1 - \ep) \Gamma (1 + 2\ep)}{\Gamma^2 (1 + \ep) \Gamma (1 - 2 \ep)}
\, .
\end{align}
Its Laurent expansion coincides with the result obtained from the parametric integral to $O (\ep^3)$. Comparing it with the
product of the one-loop expressions, we conclude that
\begin{align}
T_{21}^{\rm c1\text{-}us} = \ft14 z_1 (\ep) T_{11}^{\rm c} T_{11}^{\rm us} 
\, ,
\qquad\mbox{with}\qquad
z_1 (\ep) = \frac{\Gamma (1 + 2 \ep)}{\Gamma^2 (1 + \ep)}
\, .
\end{align}
This implies that by accounting for the mismatch factor $z_1$ in \re{cUSbrokenEq}, it becomes
\begin{align}
\label{AdjustedT2}
\sum_{n = 1}^2 \alpha_{2n} T_{2n}^{\rm c\text{-}us} = 4 z_1 (\ep) T_{11}^{\rm c} T_{11}^{\rm us} 
\, .
\end{align}
such that the multiplicative separation of the collinear and ultrasoft modes \re{MultiplicativeFactEq} can be mended by twisting the 
mode functions, $(h,J,S) \to (h/\sqrt{z_1}, \sqrt{z_1} J, \sqrt{z_1} S)$ along with a finite renormalization of the 't Hooft coupling 
\cite{Belitsky:2024yag} to this perturbative order. However, this does not seem to be possible already at three loops.

\subsection{Broken ultrasoft-collinear factorization}

Going to three loops, first, we need to re-inspect Eq.\ \re{hcUS} since it contains a mixed two-loop region, which is not a part of the
incoherent components \re{hEq}, \re{JEq}, and \re{SEq}. But, we already know that it will be broken when $T_{2n}^{\rm c\text{-}us}$
is naively recast in terms of the product $T_{11}^{\rm c} T_{11}^{\rm us}$ as was observed in Eq.\ \re{cUSbrokenEq}. To save the
day, it needs an adjustment factor. Indeed, it is the very same $z^2_1$ as in Eq.\ \re{AdjustedT2}, so that Eq.\ \re{hcUS} is superseded by
\begin{align}
\sum_{n = 1}^8 \alpha_{3n} T_{3n}^{\rm h\text{-}c\text{-}us} = - 8 z_1 (\ep) T_{11}^{\rm h} T_{11}^{\rm c}  T_{11}^{\rm us}
\, .
\end{align}
This is the harbinger of a problem. Now, the twisting implemented at two loops does not work since it produces a single factor of
$\sqrt{z_1}$ rather than the required $z_1$ in the simple, multiplicative factorization. Though replacing the product with a convolution 
could reproduce it, we will not dwell on it since further considerations of the remaining three-loop mixed regions involving only collinear 
and ultrasoft modes introduce irreparable complications in our attempt to separate them.

An inspection of both ${\rm c\text{-}c\text{-}us}$ and ${\rm c\text{-}us\text{-}us}$ regions demonstrates that they cannot simply be 
represented in terms of lower-loop incoherent components; instead, they require additional adjustment factors. With a little ingenuity,
they can be found. The equations 
\begin{align}
&
\sum_{n = 1}^8 \alpha_{3n} T_{3n}^{\rm c\text{-}c\text{-}us} 
= 
- 2  z^2_1 (\ep) z^\prime_{21} (\ep) T_{11}^{\rm us} \sum_{n = 1}^2 \alpha_{2n} T_{2n}^{\rm c\text{-}c}
\, , \\
&
\sum_{n = 1}^8 \alpha_{3n} T_{3n}^{\rm c\text{-}us\text{-}us} 
= 
- 2  z^2_1 (\ep) z^\prime_{22} (\ep) T_{11}^{\rm c} \sum_{n = 1}^2 \alpha_{2n} T_{2n}^{\rm us\text{-}us}
\, ,
\end{align}
hold to $O (\ep^0)$ for
\begin{align}
z^\prime_{21} (\ep) 
&= 1 + \ft12 \zeta_4 \ep^4 + \left( \ft72 \zeta_5 - \ft{19}4 \zeta_2 \zeta_3 \right) \ep^5 
+ 
\left( \ft{2819}{192} \zeta_6 + \ft{65}8 \zeta_3^2 \right) \ep^6 + O (\ep^7) 
\, , \\
z^\prime_{22} (\ep) 
&= 1 + 4 \zeta_4 \ep^4 - \left( 2 \zeta_5 + 8 \zeta_2 \zeta_3 \right) \ep^5 
+ 
\left( \ft{2617}{48} \zeta_6 + \ft{13}2 \zeta_3^2 \right) \ep^6 + O (\ep^7) 
\, .
\end{align}
Notice that the factorization breaking is postponed to order $\ep^4$ in the $z_{2n}$-factors when $z^2_1$, determined
at earlier loop order, is included. Unfortunately, $z_{21}$ and $z_{22}$ are different, which prevents us from twisting the 
two-loop incoherent components of $J$ and $S$ to accommodate this violation.

The inability to separate the collinear and ultrasoft regions necessitates a modification in Eq.\ \re{MultiplicativeFactEq}
with a factorization-restoring factor $\mathcal{R}$
\begin{align}
J \big( \ep, \sqrt{t} \big) S (\ep, t) \to J \big( \ep, \sqrt{t} \big) S (\ep, t) \mathcal{R} (\ep, t)
\, ,
\end{align}
with 
\begin{align}
\mathcal{R} (\ep, t) 
&
= 
1 
+ 
4 g^4 \left( z_1 - 1 \right) t^{-3 \ep} T_{11}^{\rm c} T_{11}^{\rm us}
\\
&
-
2 g^6 
\bigg[ 
\left( z_{21} - 1 \right) t^{-4 \ep} T_{11}^{\rm us} \sum_{n=1}^2 \alpha_{2n} T_{2n}^{\rm c\text{-}c} 
+
\left( z_{22} - 1 \right) t^{-5 \ep} T_{11}^{\rm c} \sum_{n=1}^2 \alpha_{2n} T_{2n}^{\rm us\text{-}us} 
\bigg]
+
O (g^6)
\, , \nonumber
\end{align}
and the $z_{2n}$-factors being
\begin{align}
z_{21} (\ep) 
&= 1 + \ft{17}{4} \zeta_4 \ep^4 + \left( \ft72 \zeta_5 - \ft{49}4 \zeta_2 \zeta_3 \right) \ep^5 
+ 
\left( \ft{283}{6} \zeta_6 + \ft{137}8 \zeta_3^2 \right) \ep^6 + O (\ep^7) 
\, , \\
z_{22} (\ep) 
&= 1 + 19 \zeta_4 \ep^4 - \left( 2 \zeta_5 + 29 \zeta_2 \zeta_3 \right) \ep^5 
+ 
\left( \ft{6217}{48} \zeta_6 + \ft{49}2 \zeta_3^2 \right) \ep^6 + O (\ep^7) 
\, .
\end{align}
Considering the fact that at each $\ell$-th order of the perturbation theory, the Laurent series possesses the leading singularity
$O (\ep^{- 2 \ell})$, the breaking of the ultrasoft-collinear factorization is pushed to the subleading order $O (\ep^{- 2})$. This 
is the order at which the violation first occurred at two loops, as was clarified in Sect.\ \ref{UScolSect}.

\section{Conclusions}

In this work, we considered factorization properties of the off-shell Sudakov form factor at higher orders of perturbative expansion.
We observed that the hard factorization is valid, i.e., short-distance physics is indeed incoherent from the IR physics governing its
near mass-shell behavior. However, we found a violation of the IR factorization for the collinear and ultrasoft modes. These remain 
entangled. It starts at two loops in subleading orders of the $\ep$-expansion. It is unsettling that an IR-sensitive observable as 
simple as the Sudakov form factor defies traditional soft-collinear factorization. This appears to suggest that all on-shell scattering 
amplitudes on this Coulomb branch will follow suit. In this paper, we did not address the question of the operator definition for the 
joint ultrasoft-collinear functions. This could be of value, and our findings will be reported elsewhere.

\begin{acknowledgments}
We are deeply indebted to Giulio Salvatori for a fruitful collaboration on this paper, which would not have been possible without his 
contribution. The implementation of his method was crucial \cite{Salvatori:2024nva} for the completion of this project. We are grateful 
to Andrei Pikelner for providing details of the calculation reported in Ref.\ \cite{Pikelner:2021goo}. The work of A.B. was supported by 
the U.S.\ National Science Foundation under grant No.\ PHY-2207138. The work of V.S. was conducted under the state assignment 
of Lomonosov Moscow State University and supported by the Moscow Center for Fundamental and Applied Mathematics of Lomonosov 
Moscow State University under Agreement No.\ 075-15-2025-345.
\end{acknowledgments}

\appendix

\appendix
\section{$\ep$-expansion for regions}

In this appendix, we present Laurent expansions for regions at one, two, and three loops. Whenever there is an exact result in
$\ep$, we quote it instead.

\subsection{One loop}
\label{Appendix1L}

At one-loop, contributions of all regions can be evaluated as exact functions of $\ep$:
\begin{align}
T_{11}^{\rm h}
&=
\e^{\ep \gamma_{\rm\scriptscriptstyle E}} \frac{\Gamma^2 (- \ep) \Gamma (1 + \ep)}{\Gamma (1 - 2 \ep)}
\, , \\
T_{11}^{\rm c1} 
&
= 
T_{11}^{\rm c2}
=
\e^{\ep \gamma_{\rm\scriptscriptstyle E}} \frac{\Gamma^2 (- \ep) \Gamma (\ep)}{2 \Gamma (- 2 \ep)}
\, , \\
T_{11}^{\rm us}
&
=
\e^{\ep \gamma_{\rm\scriptscriptstyle E}} \Gamma (1 - \ep) \Gamma (\ep)^2
\, .
\end{align}

\subsection{Two loops}
\label{Appendix2L}

At two loops, the results are required to order $O(\ep^2)$ compared to $O (\ep^0)$ of Ref.\ \cite{Belitsky:2024yag}. They are 
\begin{align}
T_{21}^{\rm h\text{-}h}
&=
\frac{1}{4 \ep^4}
+\frac{5 \pi^2}{24 \ep^2}
+
\frac{29 \zeta_3}{6 \ep}
+
\frac{3 \pi^4}{32}
+
\ep \left(\frac{329 \zeta_5}{10}-\frac{107 \pi ^2 \zeta_3}{36}\right)
+
\ep^2 \left(\frac{1723 \pi ^6}{60480} -\frac{833 \zeta_3^2}{18}\right)
\, , \\
T_{21}^{\rm h\text{-}c}
&=
-
\frac{1}{\ep^4}
-
\frac{\pi ^2}{3 \ep^2}
-
\frac{22 \zeta_3}{3 \ep}
-
\frac{7 \pi ^4}{30}
+
\ep \left(\frac{50\pi ^2 \zeta_3}{9}-\frac{478 \zeta_5}{5}\right)
+
\ep^2 \left(\frac{622 \zeta_3^2}{9}-\frac{113 \pi ^6}{756}\right)
\, , \\
T_{21}^{\rm h\text{-}us}
&=
\frac{\e^{2 \ep \gamma_{\rm E}} \Gamma (1-\ep) \Gamma (-\ep)^2 \Gamma (\ep)^2 \Gamma (1+\ep)}{\Gamma (1-2 \ep)}
\, , \\
T_{21}^{\rm c\text{-}c}
&=
\frac{1}{2 \ep^4}
+
\frac{\pi ^2}{12 \ep^2}
+
\frac{5 \zeta_3}{3 \ep}
+
\frac{163 \pi ^4}{720}
+
\ep \left(\frac{329 \zeta_5}{5}-\frac{13 \pi ^2 \zeta_3}{18}\right)
+
\ep^2 \left(\frac{6451 \pi ^6}{30240}-\frac{164 \zeta_3^2}{9}\right)
\, , \\
T_{21}^{\rm c\text{-}us}
&=
-\frac{1}{\ep^4}
-
\frac{\pi ^2}{3 \ep^2}
+
\frac{14 \zeta_3}{3 \ep}
-
\frac{\pi ^4}{15}
+
\ep \left(\frac{14 \pi^2 \zeta_3}{9}+\frac{62 \zeta_5}{5}\right)
+
\ep^2 \left(-\frac{98 \zeta_3^2}{9}-\frac{2 \pi ^6}{189}\right)
\, , \\
T_{21}^{\rm us\text{-}us}
&=
\e^{2 \ep \gamma_{\rm E}} \Gamma (-\ep)^2 \Gamma (2 \ep)^2
\, , 
\end{align}
and
\begin{align}
T_{22}^{\rm h\text{-}h}
&=
\frac{1}{\ep^4}
-
\frac{\pi ^2}{\ep^2}
-
\frac{83 \zeta_3}{3 \ep}
-
\frac{59 \pi ^4}{120}
+
\ep\left(\frac{79 \pi^2 \zeta_3}{6}-\frac{587 \zeta_5}{5}\right)
+
\ep^2 \left(\frac{2567 \zeta_3^2}{9}+\frac{59 \pi ^6}{1512}\right)
\, , \\
T_{22}^{\rm h\text{-}c}
&=
-
\frac{4}{\ep^4}
+
\frac{200 \zeta_3}{3 \ep}
+
\frac{7 \pi ^4}{5}
+
\frac{8 \pi ^2}{3 \ep^2}
+
\ep\left(\frac{2408 \zeta_5}{5}-\frac{256 \pi ^2 \zeta_3}{9}\right)
+
\ep^2 \left(\frac{257 \pi ^6}{378}-\frac{3272 \zeta_3^2}{9}\right)
\, , \\
T_{22}^{\rm h\text{-}us}
&= 0
\, , \\
T_{22}^{\rm c\text{-}c}
&=
\frac{6}{\ep^4}
-
\frac{54 \zeta_3}{\ep}
-
\frac{209 \pi ^4}{180}
-
\frac{4 \pi ^2}{3 \ep^2}
+
\ep\left(\frac{89 \pi ^2 \zeta_3}{9}-\frac{1722 \zeta_5}{5}\right)
+
\ep^2 \left(\frac{752 \zeta_3^2}{3}-\frac{185 \pi ^6}{252}\right)
\, , \\
T_{22}^{\rm c\text{-}us}
&=
-
\frac{4}{\ep^4}
+
\frac{56 \zeta_3}{3 \ep}
-
\frac{4 \pi ^4}{15}
-
\frac{4 \pi ^2}{3 \ep^2}
+
\ep\left(\frac{56 \pi ^2 \zeta_3}{9}+\frac{248 \zeta_5}{5}\right)
+
\ep^2 \left(-\frac{392 \zeta_3^2}{9}-\frac{8 \pi ^6}{189}\right)
\, , \\
T_{22}^{\rm us\text{-}us}
&=
2 \e^{2 \ep \gamma_{\rm E}} \Gamma (-\ep) \Gamma (2 \ep)^2 \big[ \Gamma (-\ep)+2 \Gamma (-2 \ep) \Gamma (1 + \ep)\big]
\, .
\end{align}

\subsection{Three loops}
\label{Appendix3L}

Finally, at three loops, the expansion was obtained to $O(\ep^0)$:
\begin{align}
T_{31}^{\rm h\text{-}h\text{-}h}
&
=
\frac{1}{36 \ep ^6}
+
\frac{11 \pi^2}{144 \ep ^4}
+
\frac{61 \zeta_3}{36 \ep ^3}+\frac{497 \pi ^4}{5760 \ep ^2}
+
\frac{1}{\ep}\left( \frac{47 \pi ^2 \zeta_3}{144}+\frac{1469\zeta_5}{60} \right)
+
\frac{49 \zeta_3^2}{72}
+
\frac{36331 \pi^6}{622080} 
\, , \\
T_{31}^{\rm h\text{-}h\text{-}c}
&
=
-
\frac{1}{6\ep ^6}-\frac{7 \pi ^2}{24 \ep ^4}
-
\frac{13 \zeta_3}{2 \ep ^3}-\frac{127 \pi ^4}{576 \ep^2}
+
\frac{1}{\ep} \left( \frac{47 \pi ^2\zeta_3}{24}-\frac{679 \zeta_5}{10} \right)
+
\frac{149 \zeta_3^2}{4}-\frac{103561 \pi ^6}{725760}
\, , \\
T_{31}^{\rm h\text{-}c\text{-}c}
&
=
\frac{1}{6 \ep^6}
+
\frac{13 \pi ^2}{72 \ep^4}
+
\frac{23 \zeta_3}{6 \ep^3}
+
\frac{443 \pi ^4}{2880 \ep^2}
+
\frac{1}{\ep}\left( \frac{519 \zeta_5}{10}-\frac{205 \pi ^2 \zeta_3}{72} \right)
-
\frac{539 \zeta_3^2}{12} + \frac{86657 \pi ^6}{725760}
\, , \\
T_{31}^{\rm h\text{-}h\text{-}us}
&
=
\frac{1}{4 \ep^6}
+
\frac{13 \pi ^2}{48 \ep^4}
+
\frac{19 \zeta_3}{4 \ep^3}
+
\frac{299 \pi ^4}{1920 \ep^2}
+
\frac{1}{\ep} \left( \frac{657 \zeta_5}{20}-\frac{89 \pi ^2 \zeta_3}{48} \right)
-
\frac{383 \zeta_3^2}{8}
+
\frac{661 \pi ^6}{10752}
\, , \\
T_{31}^{\rm c\text{-}c\text{-}c}
&
=
-
\frac{1}{18 \ep^6}
-
\frac{\pi ^2}{24 \ep^4}
-
\frac{13 \zeta_3}{18 \ep^3}
-
\frac{659 \pi ^4}{8640 \ep^2}
+
\frac{1}{\ep} \left( \frac{19 \pi ^2 \zeta_3}{24}-\frac{143 \zeta_5}{10} \right)
+
\frac{839 \zeta_3^2}{36}
-
\frac{93301 \pi ^6}{2177280}
\, , \\
T_{31}^{\rm h\text{-}c\text{-}us}
&
=
-
\frac{1}{2 \ep^6}
-
\frac{3 \pi ^2}{8 \ep^4}
-
\frac{5 \zeta_3}{2 \ep^3}
-
\frac{81 \pi ^4}{320 \ep^2}
+
\frac{1}{\ep} \left( \frac{17 \pi ^2 \zeta_3}{8}-\frac{447 \zeta_5}{10} \right)
+
\frac{167 \zeta_3^2}{4}
-
\frac{40469 \pi ^6}{241920}
\, , \\
T_{31}^{\rm c\text{-}c\text{-}us}
&
=
\frac{1}{6 \ep^6}
+
\frac{\pi ^2}{8 \ep^4}
-
\frac{\zeta_3}{2 \ep^3}
+
\frac{29 \pi ^4}{192 \ep^2}
+
\frac{1}{\ep} \left( \frac{149 \zeta_5}{10}-\frac{17 \pi ^2 \zeta_3}{24} \right)
-
\frac{25 \zeta_3^2}{4}
+
\frac{17443 \pi ^6}{103680}
\, , \\
T_{31}^{\rm h\text{-}us\text{-}us}
&
=
\frac{\Gamma (1-\ep )^2 \Gamma (-\ep )^2 \Gamma (\ep )^2 \Gamma (2 \ep
   )^2}{\Gamma (1-2 \ep ) \Gamma (\ep +1)}
\, , \\
T_{31}^{\rm c\text{-}us\text{-}us}
&
=
\frac{2 \Gamma (1-\ep )^3 \Gamma (-\ep ) \Gamma (\ep )^2 \Gamma (2 \ep
   )^2 \Gamma (3 \ep )}{\Gamma (1-2 \ep ) \Gamma (\ep +1)^2 \Gamma (2
   \ep +1)}
\, ,\\
T_{31}^{\rm us\text{-}us\text{-}us}
&
=
\frac{\Gamma (1-\ep )^3 \Gamma (\ep )^2 \Gamma (2 \ep )^2 \Gamma (3 \ep
   )^2}{\Gamma (\ep +1)^2 \Gamma (2 \ep +1)^2}
\, ,
\end{align}
for $T_{31}$
\begin{align}
T_{32}^{\rm h\text{-}h\text{-}h}
&
=
\frac{1}{36 \ep^6}
+
\frac{7 \pi ^2}{144 \ep^4}
+
\frac{55 \zeta_3}{36 \ep^3}
+
\frac{5329 \pi ^4}{51840 \ep^2}
+
\frac{1}{\ep} \left( \frac{1171 \pi^2 \zeta_3}{432}+\frac{1199 \zeta_5}{60} \right)
\\
&\qquad\qquad\qquad\qquad\qquad\qquad\qquad\qquad\qquad\qquad\qquad\quad
+
\frac{353 \zeta_3^2}{8}
+
\frac{2606843 \pi ^6}{13063680}
\, , \nonumber\\
T_{32}^{\rm h\text{-}h\text{-}c}
&
=
-
\frac{1}{6 \ep^6}
-
\frac{13 \pi ^2}{72 \ep^4}
-
\frac{35 \zeta_3}{6 \ep^3}
-
\frac{2561 \pi ^4}{8640 \ep^2}
-
\frac{1}{\ep} \left( \frac{247 \pi ^2 \zeta_3}{72} + \frac{739 \zeta_5}{10} \right)
-
\frac{497 \zeta_3^2}{12}
-
\frac{306353 \pi ^6}{725760}
\, , \\
T_{32}^{\rm h\text{-}c\text{-}c}
&
=
\frac{5}{12 \ep^6}
+
\frac{5 \pi ^2}{16 \ep^4}
+
\frac{31 \zeta_3}{4 \ep^3}
+
\frac{6541 \pi ^4}{17280 \ep^2}
+
\frac{1}{\ep} \left( \frac{417 \zeta_5}{4} - \frac{107 \pi ^2 \zeta_3}{144} \right)
-
\frac{979 \zeta_3^2}{24}
+
\frac{107749 \pi ^6}{290304}
\, , \\
T_{32}^{\rm h\text{-}h\text{-}us}
&
= 0
\, , \\
T_{32}^{\rm c\text{-}c\text{-}c}
&
=
-
\frac{1}{18 \ep^6}
-
\frac{\pi ^2}{24 \ep^4}
-
\frac{13 \zeta_3}{18 \ep^3}
-
\frac{1657 \pi ^4}{25920 \ep^2}
+
\frac{1}{\ep} \left( \frac{59 \pi^2 \zeta_3}{216} - \frac{829 \zeta_5}{30} \right)
-
\frac{25 \zeta_3^2}{4}
-
\frac{42857 \pi ^6}{933120}
\, , \\
T_{32}^{\rm h\text{-}c\text{-}us}
&
=
-
\frac{1}{2 \ep^6}
-
\frac{3 \pi ^2}{8 \ep^4}
-
\frac{5 \zeta_3}{2 \ep^3}
-
\frac{81 \pi ^4}{320 \ep^2}
+
\frac{1}{\ep} \left( \frac{17 \pi^2 \zeta_3}{8} - \frac{447 \zeta_5}{10} \right)
+
\frac{167 \zeta_3^2}{4}
-
\frac{40469 \pi ^6}{241920}
\, , \\
T_{32}^{\rm c\text{-}c\text{-}us}
&
=
\frac{1}{6 \ep^6}
+
\frac{13 \pi ^2}{72 \ep^4}
-
\frac{\zeta_3}{6 \ep^3}
+
\frac{1073 \pi ^4}{8640 \ep^2}
+
\frac{1}{\ep}
\left( \frac{239 \zeta_5}{10}
-
\frac{125 \pi ^2 \zeta_3}{72} \right)
+
\frac{65 \zeta_3^2}{12}
+
\frac{70433 \pi ^6}{725760}
\, , \\
T_{32}^{\rm h\text{-}us\text{-}us}
&
=
\frac{1}{4 \ep^6}
+
\frac{13 \pi ^2}{48 \ep^4}
-
\frac{5 \zeta_3}{4 \ep^3}
+
\frac{331 \pi ^4}{1920 \ep^2}
-
\frac{1}{\ep} \left( \frac{65 \pi ^2 \zeta_3}{48} + \frac{63 \zeta_5}{20} \right)
+
\frac{25 \zeta_3^2}{8}
+
\frac{1997 \pi ^6}{23040}
\, , \\
T_{32}^{\rm c\text{-}us\text{-}us}
&
=
-
\frac{1}{6 \ep^6}
-
\frac{7 \pi ^2}{24 \ep^4}
+
\frac{3 \zeta_3}{2 \ep^3}
-
\frac{2513 \pi ^4}{8640 \ep^2}
+
\frac{1}{\ep} \left( \frac{205 \pi ^2 \zeta_3}{72} + \frac{81 \zeta_5}{10} \right)
-
\frac{73 \zeta_3^2}{12}
-
\frac{4931 \pi ^6}{20736}
\, ,\\
T_{32}^{\rm us\text{-}us\text{-}us}
&
=
\frac{1}{36 \ep^6}
+
\frac{11 \pi ^2}{144 \ep^4}
-
\frac{11 \zeta_3}{36 \ep^3}
+
\frac{6553 \pi ^4}{51840 \ep^2}
-
\frac{1}{\ep} \left( \frac{299 \pi^2 \zeta_3}{432} + \frac{137 \zeta_5}{20} \right)
+
\frac{\zeta_3^2}{8}
+
\frac{2086247 \pi ^6}{13063680}
\, ,
\end{align}
for $T_{32}$
\begin{align}
T_{33}^{\rm h\text{-}h\text{-}h}
&
=
-
\frac{1}{36 \ep^6}
+
\frac{17 \pi ^2}{144 \ep^4}
+
\frac{149 \zeta_3}{36 \ep^3}
+
\frac{3701 \pi ^4}{17280 \ep^2}
+
\frac{1}{\ep} \left( \frac{967 \zeta_5}{20} - \frac{229 \pi ^2 \zeta_3}{144} \right)
\\
&\qquad\qquad\qquad\qquad\qquad\qquad\qquad\qquad\qquad\qquad\qquad
-
\frac{11113 \zeta_3^2}{72}
+
\frac{367151 \pi ^6}{4354560}
\, , \nonumber\\
T_{33}^{\rm h\text{-}h\text{-}c}
&
=
\frac{1}{6 \ep^6}
-
\frac{35 \pi ^2}{72 \ep^4}
-
\frac{97 \zeta_3}{6 \ep^3}
-
\frac{5743 \pi ^4}{8640 \ep^2}
+
\frac{1}{\ep} \left( \frac{595 \pi ^2  \zeta_3}{72}-\frac{2121 \zeta_5}{10} \right)
+
\frac{3905 \zeta_3^2}{12}
-
\frac{43447 \pi ^6}{80640}
\, , \\
T_{33}^{\rm h\text{-}c\text{-}c}
&
=
-
\frac{1}{6 \ep^6}
+
\frac{23 \pi ^2}{72 \ep^4}
+
\frac{79 \zeta_3}{6 \ep^3}
+
\frac{1661 \pi ^4}{2880 \ep^2}
+
\frac{1}{\ep}
\left( \frac{2551 \zeta_5}{10}
-
\frac{521 \pi ^2 \zeta_3}{72} \right)
-
\frac{2089 \zeta_3^2}{12}
+
\frac{88013 \pi ^6}{103680}
\, , \\
T_{33}^{\rm h\text{-}h\text{-}us}
&
=
-
\frac{1}{4 \ep^6}
+
\frac{19 \pi ^2}{48 \ep^4}
+
\frac{45 \zeta_3}{4 \ep^3}
+
\frac{199 \pi ^4}{640 \ep^2}
+
\frac{1}{\ep}
\left( \frac{903 \zeta_5}{20}
-
\frac{167 \pi ^2 \zeta_3}{48} \right)
-
\frac{1057 \zeta_3^2}{8}
+
\frac{12911 \pi ^6}{483840}
\, , \\
T_{33}^{\rm c\text{-}c\text{-}c}
&
=
\frac{1}{18 \ep^6}
-
\frac{5 \pi ^2}{72 \ep^4}
-
\frac{83 \zeta_3}{18 \ep^3}
-
\frac{1709 \pi ^4}{8640 \ep^2}
+
\frac{1}{\ep} \left( \frac{127 \pi ^2 \zeta_3}{72}-\frac{2491 \zeta_5}{30} \right)
+
\frac{3001 \zeta_3^2}{36}
-
\frac{637379 \pi ^6}{2177280}
\, , \\
T_{33}^{\rm h\text{-}c\text{-}us}
&
=
\frac{1}{2 \ep^6}
-
\frac{7 \pi ^2}{24 \ep^4}
-
\frac{27 \zeta_3}{2 \ep^3}
-
\frac{397 \pi ^4}{960 \ep^2}
+
\frac{1}{\ep} \left( \frac{29 \pi ^2 \zeta_33}{24}-\frac{993 \zeta_5}{10} \right)
+
\frac{409 \zeta_3^2}{4}
-
\frac{71867 \pi ^6}{241920}
\, , \\
T_{33}^{\rm c\text{-}c\text{-}us}
&
=
-
\frac{5}{12 \ep^6}
-
\frac{17 \pi ^2}{144 \ep^4}
+
\frac{89 \zeta_3}{12 \ep^3}
+
\frac{319 \pi ^4}{3456 \ep^2}
+
\frac{1}{\ep} \left( \frac{373 \pi^2 \zeta_3}{144}+\frac{213 \zeta_5}{4} \right)
-
\frac{1549 \zeta_3^2}{24}
+
\frac{32621 \pi ^6}{207360}
\, , \\
T_{33}^{\rm h\text{-}us\text{-}us}
&
= 0
\, , \\
T_{33}^{\rm c\text{-}us\text{-}us}
&
=
-\frac{2 \Gamma (1-\ep )^3 \Gamma (-\ep ) \Gamma (\ep )^2 \Gamma (2 \ep
   )^2 \Gamma (3 \ep )}{\Gamma (1-2 \ep ) \Gamma (\ep +1)^2 \Gamma (2
   \ep +1)}
\, ,\\
T_{33}^{\rm us\text{-}us\text{-}us}
&
=
-\frac{\Gamma (1-\ep )^3 \Gamma (\ep )^2 \Gamma (2 \ep )^2 \Gamma (3
   \ep )^2}{\Gamma (\ep +1)^2 \Gamma (2 \ep +1)^2}
\, ,
\end{align}
for $T_{33}$
\begin{align}
T_{34}^{\rm h\text{-}h\text{-}h}
&
=
-
\frac{1}{36 \ep^6}
-
\frac{11 \pi ^2}{144 \ep^4}
-
\frac{109 \zeta_3}{36 \ep^3}
-
\frac{59 \pi ^4}{1920 \ep^2}
+
\frac{1}{\ep}
\left( \frac{1153 \pi^2 \zeta_3}{144} - \frac{5989 \zeta_5}{60} \right)
\\
&\qquad\qquad\qquad\qquad\qquad\qquad\qquad\qquad\qquad\qquad\qquad\quad
-
\frac{1009 \zeta_3^2}{72}
-
\frac{1155757 \pi ^6}{4354560}
\, , \nonumber\\
T_{34}^{\rm h\text{-}h\text{-}c}
&
=
\frac{1}{6 \ep^6}
+
\frac{7 \pi ^2}{24 \ep^4}
+
\frac{21 \zeta_3}{2 \ep^3}
+
\frac{731 \pi ^4}{2880 \ep^2}
+
\frac{1}{\ep} \left( \frac{2799 \zeta_5}{10} - \frac{455 \pi ^2 \zeta_3}{24} \right)
-
\frac{501 \zeta_3^2}{4}
+
\frac{469033 \pi ^6}{725760}
\, , \\
T_{34}^{\rm h\text{-}c\text{-}c}
&
=
-
\frac{5}{12 \ep^6}
-
\frac{65 \pi ^2}{144 \ep^4}
-
\frac{151 \zeta_3}{12 \ep^3}
-
\frac{3127 \pi ^4}{5760 \ep^2}
+
\frac{1}{\ep} \left( \frac{1973  \pi ^2 \zeta_3}{144}-\frac{1211 \zeta_5}{4} \right)
\\
&\qquad\qquad\qquad\qquad\qquad\qquad\qquad\qquad\qquad\qquad\qquad\quad
+
\frac{4915 \zeta_3^2}{24}
-
\frac{34727 \pi ^6}{41472}
\, , \nonumber\\
T_{34}^{\rm h\text{-}h\text{-}us}
&
=
0
\, , \\
T_{34}^{\rm c\text{-}c\text{-}c}
&
=
\frac{1}{18 \ep^6}
+
\frac{\pi ^2}{24 \ep^4}
+
\frac{37 \zeta_3}{18 \ep^3}
+
\frac{1907 \pi ^4}{8640 \ep^2}
+
\frac{1}{\ep} \left( \frac{2869 \zeta_5}{30} - \frac{9 \pi ^2 \zeta_3}{8} \right)
-
\frac{1799 \zeta_3^2}{36}
+
\frac{135811 \pi ^6}{311040}
\, , \\
T_{34}^{\rm h\text{-}c\text{-}us}
&
=
\frac{1}{2 \ep^6}
+
\frac{3 \pi ^2}{8 \ep^4}
+
\frac{5 \zeta_3}{2 \ep^3}
+
\frac{81 \pi ^4}{320 \ep^2}
+
\frac{1}{\ep}
\left( \frac{447 \zeta_5}{10}-\frac{17 \pi ^2 \zeta_3}{8} \right)
-
\frac{167 \zeta_3^2}{4}
+
\frac{40469 \pi ^6}{241920}
\, , \\
T_{34}^{\rm c\text{-}c\text{-}us}
&
=
-
\frac{1}{6 \ep^6}
-
\frac{\pi ^2}{8 \ep^4}
+
\frac{\zeta_3}{2 \ep^3}
-
\frac{29 \pi ^4}{192 \ep^2}
+
\frac{1}{\ep}
\left(
\frac{17 \pi ^2 \zeta_3}{24}-\frac{149 \zeta_5}{10}
\right)
+
\frac{25 \zeta_3^2}{4}
-
\frac{17443 \pi ^6}{103680}
\, , \\
T_{34}^{\rm h\text{-}us\text{-}us}
&
=
-
\frac{1}{4 \ep^6}
-
\frac{3 \pi ^2}{16 \ep^4}
+
\frac{7 \zeta_3}{4 \ep^3}
-
\frac{163 \pi ^4}{1920 \ep^2}
+
\frac{1}{\ep} \left( \frac{21 \pi ^2 \zeta_3}{16}+\frac{93 \zeta_5}{20} \right)
-
\frac{49 \zeta_3^2}{8}
-
\frac{15269 \pi ^6}{483840}
\, , \\
T_{34}^{\rm c\text{-}us\text{-}us}
&
=
-\frac{2 \Gamma (1-\ep )^3 \Gamma (-\ep ) \Gamma (\ep )^2 \Gamma (2 \ep
   )^2 \Gamma (3 \ep )}{\Gamma (1-2 \ep ) \Gamma (\ep +1)^2 \Gamma (2
   \ep +1)}
\, ,\\
T_{34}^{\rm us\text{-}us\text{-}us}
&
= 
-\frac{\Gamma (1-\ep )^3 \Gamma (\ep )^2 \Gamma (2 \ep )^2 \Gamma (3
   \ep )^2}{\Gamma (\ep +1)^2 \Gamma (2 \ep +1)^2}
\, , \nonumber
\end{align}
for $T_{34}$
\begin{align}
T_{3,56}^{\rm h\text{-}h\text{-}h}
&
=
\frac{11}{36 \ep^6}
-
\frac{43 \pi ^2}{144 \ep^4}
-
\frac{307 \zeta_3}{36 \ep^3}
-
\frac{523 \pi ^4}{3456 \ep^2}
+
\frac{1}{\ep} \left( \frac{1483 \pi^2 \zeta_3}{144}-\frac{3257 \zeta_5}{20} \right)
\\
&\qquad\qquad\qquad\qquad\qquad\qquad\qquad\qquad\qquad\qquad\qquad\quad
-
\frac{6193 \zeta_3^2}{72}
-
\frac{1576517 \pi ^6}{4354560}
\, , \nonumber\\
T_{3,56}^{\rm h\text{-}h\text{-}c}
&
=
-
\frac{11}{6 \ep^6}
+
\frac{113 \pi ^2}{72 \ep^4}
+
\frac{235 \zeta_3}{6 \ep^3}
+
\frac{929 \pi ^4}{1728 \ep^2}
+
\frac{1}{\ep} \left( \frac{3991 \zeta_5}{10}-\frac{2545 \pi ^2 \zeta_3}{72} \right)
\\
&\qquad\qquad\qquad\qquad\qquad\qquad\qquad\qquad\qquad\qquad\qquad\quad
-
\frac{2315 \zeta_3^2}{12}
+
\frac{328597 \pi ^6}{725760}
\, , \nonumber\\
T_{3,56}^{\rm h\text{-}c\text{-}c}
&
=
\frac{55}{12 \ep^6}
-
\frac{127 \pi ^2}{48 \ep^4}
-
\frac{301 \zeta_3}{4 \ep^3}
-
\frac{17593 \pi ^4}{17280 \ep^2}
+
\frac{1}{\ep} \left( \frac{6245  \pi ^2 \zeta_3}{144}-\frac{1747 \zeta_5}{4} \right)
\\
&\qquad\qquad\qquad\qquad\qquad\qquad\qquad\qquad\qquad\qquad\qquad\quad
+
\frac{15211 \zeta_3^2}{24}
-
\frac{2311 \pi ^6}{10752}
\, , \nonumber\\
T_{3,56}^{\rm h\text{-}h\text{-}us}
&
= 0
\, , \\
T_{3,56}^{\rm c\text{-}c\text{-}c}
&
=
-
\frac{37}{9 \ep^6}
+
\frac{31 \pi ^2}{36 \ep^4}
+
\frac{479 \zeta_3}{9 \ep^3}
+
\frac{2177 \pi ^4}{4320 \ep^2}
+
\frac{1}{\ep} \left( \frac{859 \zeta_5}{5}-\frac{449 \pi ^2 \zeta_3}{36} \right)
-
\frac{8239 \zeta_3^2}{18}
+
\frac{2801 \pi ^6}{120960}
\, , \\
T_{3,56}^{\rm h\text{-}c\text{-}us}
&
=
-
\frac{2}{\ep^6}
+
\frac{26 \zeta_3}{\ep^3}
+
\frac{39 \pi ^4}{80 \ep^2}
+
\frac{1}{\ep} \left( \pi ^2 \zeta_3
+
\frac{726 \zeta_5}{5} \right)
-
157  \zeta_3^2
+
\frac{71 \pi ^6}{189}
\, , \\
T_{3,56}^{\rm c\text{-}c\text{-}us}
&
=
\frac{13}{3 \ep^6}
+
\frac{65 \pi ^2}{36 \ep^4}
-
\frac{146 \zeta_3}{3 \ep^3}
+
\frac{293 \pi ^4}{864 \ep^2}
-
\frac{1}{\ep} \left( \frac{205 \pi ^2 \zeta_3}{9} + \frac{768 \zeta_5}{5} \right)
+
\frac{983 \zeta_3^2}{3}
-
\frac{977 \pi ^6}{51840}
\, , \\
T_{3,56}^{\rm h\text{-}us\text{-}us}
&
=
\frac{1}{4 \ep^6}
+
\frac{13 \pi ^2}{48 \ep^4}
-
\frac{5 \zeta_3}{4 \ep^3}
+
\frac{331 \pi ^4}{1920 \ep^2}
-
\frac{1}{\ep}
\left( \frac{65 \pi ^2 \zeta_3}{48} + \frac{63 \zeta_5}{20} \right)
+
\frac{25 \zeta_3^2}{8}
+
\frac{1997 \pi ^6}{23040}
\, , \\
T_{3,56}^{\rm c\text{-}us\text{-}us}
&
=
-
\frac{11}{6 \ep^6}
-
\frac{53 \pi ^2}{24 \ep^4}
+
\frac{37 \zeta_3}{2 \ep^3}
-
\frac{14731 \pi ^4}{8640 \ep^2}
+
\frac{1}{\ep} \left( \frac{1787 \pi^2 \zeta_3}{72}+\frac{511 \zeta_5}{10} \right)
-
\frac{1271 \zeta_3^2}{12}
-
\frac{24185 \pi ^6}{20736}
\, ,\\
T_{3,56}^{\rm us\text{-}us\text{-}us}
&
=
\frac{11}{36 \ep^6}
+
\frac{31 \pi ^2}{48 \ep^4}
-
\frac{115 \zeta_3}{36 \ep^3}
+
\frac{1601 \pi ^4}{1920 \ep^2}
-
\frac{1}{\ep} \left( \frac{1085 \pi ^2 \zeta_3}{144} + \frac{217 \zeta_5}{20} \right)
\\
&\qquad\qquad\qquad\qquad\qquad\qquad\qquad\qquad\qquad\qquad\qquad\quad
+
\frac{1079 \zeta_3^2}{72}
+
\frac{3868787 \pi ^6}{4354560}
\, , \nonumber
\end{align}
for the sum $T_{3,56} = T_{35} + T_{36}$
\begin{align}
T_{37}^{\rm h\text{-}h\text{-}h}
&
=
\frac{1}{36 \ep^6}
+
\frac{13 \pi ^2}{432 \ep^4}
-
\frac{7 \zeta_3}{12 \ep^3}
+
\frac{11 \pi ^4}{17280 \ep^2}
+
\frac{1}{\ep} \left( \frac{739 \zeta_5}{60}-\frac{155 \pi ^2 \zeta_3}{144} \right)
+
\frac{275 \zeta_3^2}{24}
+
\frac{2267 \pi ^6}{1451520}
\, , \\
T_{37}^{\rm h\text{-}h\text{-}c}
&
=
-
\frac{1}{6 \ep^6}
-
\frac{5 \pi ^2}{72 \ep^4}
+
\frac{17 \zeta_3}{6 \ep^3}
+
\frac{1007 \pi ^4}{8640 \ep^2}
+
\frac{1}{\ep} \left( \frac{181 \pi ^2 \zeta_3}{72}+\frac{201 \zeta_5}{10} \right)
-
\frac{241 \zeta_3^2}{12}
+
\frac{34957 \pi ^6}{241920}
\, , \\
T_{37}^{\rm h\text{-}c\text{-}c}
&
=
\frac{1}{6 \ep^6}
+
\frac{\pi ^2}{72 \ep^4}
-
\frac{19 \zeta_3}{6 \ep^3}
-
\frac{589 \pi ^4}{2880 \ep^2}
-
\frac{1}{\ep} \left( \frac{115 \pi ^2 \zeta_3}{72} + \frac{731 \zeta_5}{10} \right)
+
\frac{73 \zeta_3^2}{12}
-
\frac{16253 \pi ^6}{48384}
\, , \\
T_{37}^{\rm h\text{-}h\text{-}us}
&
=
\frac{1}{4 \ep^6}
+
\frac{\pi ^2}{48 \ep^4}
-
\frac{11 \zeta_3}{4 \ep^3}
-
\frac{77 \pi ^4}{1920 \ep^2}
-
\frac{1}{\ep}
\left(
\frac{11 \pi ^2 \zeta_3}{48} + \frac{273 \zeta_5}{20}
\right)
+
\frac{121 \zeta_3^2}{8}
-
\frac{5233 \pi ^6}{161280}
\, , \\
T_{37}^{\rm c\text{-}c\text{-}c}
&
=
-
\frac{1}{18 \ep^6}
-
\frac{\pi ^2}{216 \ep^4}
+
\frac{3 \zeta_3}{2 \ep^3}
+
\frac{589 \pi ^4}{8640 \ep^2}
+
\frac{1}{\ep} \left( \frac{41 \pi ^2 \zeta_3}{72}+\frac{417 \zeta_5}{10} \right)
-
\frac{67 \zeta_3^2}{12}
+
\frac{460883 \pi ^6}{2177280}
\, , \\
T_{37}^{\rm h\text{-}c\text{-}us}
&
=
\frac{2 \Gamma (1-\ep )^2 \Gamma (-2 \ep ) \Gamma (-\ep )^2 \Gamma (\ep
   )^2 \Gamma (2 \ep )}{\Gamma (1-3 \ep ) \Gamma (1-2 \ep )}
\, , \\
T_{37}^{\rm c\text{-}c\text{-}us}
&
=
\frac{2 \Gamma (-2 \ep ) \Gamma (-\ep ) \Gamma (\ep )^2 \Gamma (2 \ep )
   \Gamma (3 \ep ) \Gamma (1-\ep )^3}{\Gamma (1-3 \ep ) \Gamma (1-2 \ep
   ) \Gamma (\ep +1) \Gamma (2 \ep +1)}+\frac{\Gamma (-\ep )^2 \Gamma
   (\ep )^2 \Gamma (2 \ep )^2 \Gamma (1-\ep )^3}{\Gamma (1-2 \ep )^2
   \Gamma (\ep +1)^2}
\, , \\
T_{37}^{\rm h\text{-}us\text{-}us}
&
= 0
\, , \\
T_{37}^{\rm c\text{-}us\text{-}us}
&
=
\frac{2 \Gamma (1-\ep )^3 \Gamma (-\ep ) \Gamma (\ep )^2 \Gamma (2 \ep
   )^2 \Gamma (3 \ep )}{\Gamma (1-2 \ep ) \Gamma (\ep +1)^2 \Gamma (2
   \ep +1)}
\, ,\\
T_{37}^{\rm us\text{-}us\text{-}us}
&
=
\frac{\Gamma (1-2 \ep ) \Gamma (1-\ep )^2 \Gamma (\ep )^2 \Gamma (2 \ep
   )^2 \Gamma (3 \ep )^2}{\Gamma (\ep +1) \Gamma (2 \ep +1)^2}
\, ,
\end{align}
for $T_{37}$, and finally
\begin{align}
T_{38}^{\rm h\text{-}h\text{-}h}
&
=
\frac{1}{4 \ep^6}
-
\frac{115 \pi ^2}{432 \ep^4}
-
\frac{305 \zeta_3}{36 \ep^3}
-
\frac{3149 \pi ^4}{17280 \ep^2}
+
\frac{1}{\ep} \left( \frac{1273 \pi ^2 \zeta_3}{144}-\frac{6569 \zeta_5}{60} \right)
\\
&\qquad\qquad\qquad\qquad\qquad\qquad\qquad\qquad\qquad\qquad\qquad\quad
+
\frac{9817 \zeta_3^2}{72}
-
\frac{516863 \pi ^6}{4354560}
\, , \nonumber\\
T_{38}^{\rm h\text{-}h\text{-}c}
&
=
-
\frac{3}{2 \ep^6}
+
\frac{11 \pi ^2}{8 \ep^4}
+
\frac{75 \zeta_3}{2 \ep^3}
+
\frac{197 \pi ^4}{320 \ep^2}
+
\frac{1}{\ep} \left( \frac{3609 \zeta_5}{10}-\frac{275 \pi ^2 \zeta_3}{8} \right)
-
\frac{1875 \zeta_3^2}{4}
+
\frac{7837 \pi ^6}{26880}
\, , \\
T_{38}^{\rm h\text{-}c\text{-}c}
&
=
\frac{15}{4 \ep^6}
-
\frac{331 \pi ^2}{144 \ep^4}
-
\frac{827 \zeta_3}{12 \ep^3}
-
\frac{18763 \pi ^4}{17280 \ep^2}
+
\frac{1}{\ep} \left( \frac{2095 \pi ^2 \zeta_3}{48}-\frac{1927 \zeta_5}{4} \right)
\\
&\qquad\qquad\qquad\qquad\qquad\qquad\qquad\qquad\qquad\qquad\qquad\quad
+
\frac{5503 \zeta_3^2}{8}
-
\frac{62311 \pi ^6}{161280}
\, , \nonumber\\
T_{38}^{\rm h\text{-}h\text{-}us}
&
= 0
\, , \\
T_{38}^{\rm c\text{-}c\text{-}c}
&
=
-
\frac{7}{2 \ep^6}
+
\frac{193 \pi ^2}{216 \ep^4}
+
\frac{965 \zeta_3}{18 \ep^3}
+
\frac{6131 \pi ^4}{8640 \ep^2}
+
\frac{1}{\ep} \left( \frac{8413 \zeta_5}{30}-\frac{1033 \pi ^2 \zeta_3}{72} \right)
\\
&\qquad\qquad\qquad\qquad\qquad\qquad\qquad\qquad\qquad\qquad\qquad\quad
-
\frac{17335 \zeta_3^2}{36}
+
\frac{208591 \pi ^6}{725760}
\, , \nonumber\\
T_{38}^{\rm h\text{-}c\text{-}us}
&
=
\frac{2 \Gamma (1-\ep ) \Gamma (-2 \ep ) \Gamma (-\ep )^3 \Gamma (\ep
   )^2 \Gamma (2 \ep )}{\Gamma (1-2 \ep ) \Gamma (-3 \ep )}
\, , \\
T_{38}^{\rm c\text{-}c\text{-}us}
&
=
\frac{\Gamma (1-\ep ) \Gamma (\ep )^2 \Gamma (2 \ep )^2 \Gamma (-\ep
   )^4}{\Gamma (-2 \ep )^2 \Gamma (\ep +1)^2}+\frac{2 \Gamma (1-\ep )^2
   \Gamma (\ep )^2 \Gamma (2 \ep )^2 \Gamma (-\ep )^3}{\Gamma (1-2 \ep
   ) \Gamma (-2 \ep ) \Gamma (\ep +1)^2}
 \nonumber\\ &      
   +\frac{2 \Gamma (1-\ep )^2 \Gamma
   (-2 \ep ) \Gamma (\ep )^2 \Gamma (2 \ep ) \Gamma (3 \ep ) \Gamma
   (-\ep )^2}{\Gamma (1-2 \ep ) \Gamma (-3 \ep ) \Gamma (\ep +1) \Gamma
   (2 \ep +1)}+\frac{2 \Gamma (1-\ep )^2 \Gamma (-2 \ep ) \Gamma (\ep
   )^3 \Gamma (3 \ep ) \Gamma (-\ep )^2}{\Gamma (1-2 \ep ) \Gamma (-3
   \ep ) \Gamma (\ep +1)^2}
\, , \\
T_{38}^{\rm h\text{-}us\text{-}us}
&
=
\Gamma (1-\ep ) \Gamma (-\ep )^2 \Gamma (\ep )^2 \Gamma (2 \ep )^2
\, , \\
T_{38}^{\rm c\text{-}us\text{-}us}
&
=
\frac{2 \Gamma (-\ep ) \Gamma (\ep )^2 \Gamma (2 \ep )^2 \Gamma (3 \ep
   ) \Gamma (1-\ep )^3}{\Gamma (1-2 \ep ) \Gamma (\ep +1)^2 \Gamma (2
   \ep +1)}+\frac{2 \Gamma (-\ep ) \Gamma (\ep )^3 \Gamma (2 \ep )
   \Gamma (3 \ep ) \Gamma (1-\ep )^2}{\Gamma (\ep +1)^2}
 \nonumber\\ &     
   +\frac{2 \Gamma
   (-\ep )^2 \Gamma (\ep )^3 \Gamma (2 \ep ) \Gamma (3 \ep ) \Gamma
   (1-\ep )^2}{\Gamma (-2 \ep ) \Gamma (\ep +1)^3}+\frac{2 \Gamma (1-2
   \ep ) \Gamma (-\ep )^2 \Gamma (\ep )^2 \Gamma (2 \ep )^2 \Gamma (3
   \ep ) \Gamma (1-\ep )}{\Gamma (-2 \ep ) \Gamma (\ep +1) \Gamma (2
   \ep +1)}
\, ,\\
T_{38}^{\rm us\text{-}us\text{-}us}
&
= 
\frac{\Gamma (1-2 \ep ) \Gamma (1-\ep )^2 \Gamma (3 \ep )^2 \Gamma (\ep
   )^4}{\Gamma (\ep +1)^3}+\frac{2 \Gamma (1-2 \ep ) \Gamma (1-\ep )^2
   \Gamma (2 \ep ) \Gamma (3 \ep )^2 \Gamma (\ep )^3}{\Gamma (\ep +1)^2
   \Gamma (2 \ep +1)}
 \nonumber\\ &  
   +\frac{\Gamma (1-3 \ep ) \Gamma (1-\ep )^2 \Gamma (2
   \ep )^2 \Gamma (3 \ep )^2 \Gamma (\ep )^2}{\Gamma (\ep +1)^2 \Gamma
   (2 \ep +1)}
\, , 
\end{align}
for $T_{38}$.

Mathematica-ready expressions for individual regions of the above three-loop results are given in the attached ancillary files.


\end{document}